\documentclass[a4paper]{article}

\usepackage{graphicx}
\usepackage{natbib}
\usepackage{amsmath}
\usepackage{amsfonts}
\usepackage{url}

\newcounter{ctr}

\newcounter{ctr1}

\newcounter{ctr2}

\newcounter{ctr3}

\newenvironment{theorem*}[1]{{\bf Theorem #1} \begin{itshape}}{\end{itshape}}

\newenvironment{corollary*}[1]{{\bf Corollary #1} \begin{itshape}}{\end{itshape}}

\newenvironment{proposition*}[1]{{\bf Proposition #1} \begin{itshape}}{\end{itshape}}

\newcommand{\ud}{\, {\rm d} \kern-.015em }

\newcommand{\modulus}[1]{\left| \kern.05em #1 \kern.05em \right|}
\newcommand{\norm}[1]{\left\| \kern.05em #1 \kern.05em \right\|}
\newcommand{\inner}[1]{\left\langle \kern.05em #1 \kern.05em \right\rangle }

\newcommand{\pick}[2]{\renewcommand{\arraystretch}{0.6}
\left( \kern-.4em \begin{array}{c} #1 \\ #2 \end{array} \kern-.4em \right) }

\newcommand{\z}{\mathbf{z}}
\newcommand{\Z}{\mathbf{Z}}
\newcommand{\bb}{\mathbf{b}}
\newcommand{\be}{\mathbf{e}}
\newcommand{\B}{\mathbf{B}}
\newcommand{\x}{\mathbf{x}}
\newcommand{\X}{\mathbf{X}}
\newcommand{\y}{\mathbf{y}}
\newcommand{\Y}{\mathbf{Y}}
\newcommand{\U}{\mathbf{U}}
\newcommand{\bu}{\mathbf{u}}
\newcommand{\bv}{\mathbf{v}}
\newcommand{\bV}{\mathbf{V}}

\usepackage{amsfonts}
\newcommand{\E}{\mathbb E}

\newcommand{\R}{\mathbb R}

\usepackage{colortbl}
\usepackage[ruled,vlined]{algorithm2e}
\usepackage{enumerate}
\usepackage{subfig}
\usepackage{stmaryrd}
\usepackage{natbib}

\usepackage{authblk}
\usepackage{xcolor}

\title{Piecewise Deterministic Markov Processes for Continuous-Time Monte Carlo}
\author[1,$\dag$]{Paul Fearnhead}
\author[2]{Joris Bierkens}
\author[2]{ Murray Pollock}
\author[2]{ Gareth O Roberts}
\affil[1]{Department of Mathematics and Statistics, Lancaster University}
\affil[2]{Department of Statistics, University of Warwick}
\affil[$\dag$]{Correspondence: p.fearnhead@lancaster.ac.uk}

\begin{document}
\maketitle 
\begin{center}
 {\bf Abstract} 
 \end{center} Recently there have been exciting developments in Monte Carlo methods, with the development of new
 MCMC and sequential Monte Carlo (SMC) algorithms which are based on continuous-time, rather
 than discrete-time, Markov processes. This has led to some fundamentally new Monte Carlo
 algorithms which can be used to sample from, say, a posterior distribution. 
  Interestingly, continuous-time algorithms seem particularly well suited to Bayesian analysis in big-data settings as they
 need only access a small sub-set of data points at each iteration, and yet are still guaranteed to target the
 true posterior distribution. Whilst continuous-time MCMC and SMC methods have been developed independently we show here that they
 are related by the fact that both involve simulating a
 piecewise deterministic Markov process.  Furthermore we show that the methods developed to date are just specific cases of a potentially
 much wider class of continuous-time Monte Carlo algorithms. We give an informal introduction to piecewise deterministic Markov processes, 
 covering the aspects relevant to these new Monte Carlo algorithms, with a view to making the development of new continuous-time Monte Carlo more
 accessible. We focus on how and why sub-sampling ideas can be used with these algorithms, and aim to give insight into how these new algorithms
 can be implemented, and what are some of the issues that affect their efficiency.

 \section{Introduction}
\label{sec:introduction}

Monte Carlo methods, such as MCMC and SMC, have been central to the application of Bayesian statistics to real-world problems \cite[]{Robert/Casella:2011,McGrayne:2011}. These 
established Monte Carlo methods are based upon simulating discrete-time Markov processes. For example MCMC algorithms simulate a discrete-time Markov chain
constructed to have a target distribution of interest, the posterior distribution in Bayesian inference, as its stationary distribution. Whilst SMC methods
involve propagating and re-weighting particles so that a final set of weighted particles approximate a target distribution. The propagation step here also involves
simulating from a discrete-time Markov chain.

In the past few years there have been exciting developments in MCMC and SMC methods based on continuous-time versions of these Monte Carlo methods. For example, continuous-time MCMC algorithms
have been proposed \cite[]{Peters:2012,Bouchard:2016,Bierkens/Roberts:2015,Bierkens/Fearnhead/Roberts:2016} that involve simulating a continuous-time Markov process that
has been designed to have a target distribution of interest as its stationary distribution. These continuous-time MCMC algorithms were originally motivated as they
are examples of non-reversible Markov processes. There is substantial evidence that non-reversible MCMC algorithms
will be more efficient than standard MCMC algorithms that are reversible \cite[]{neal1998suppressing,DiaconisHolmesNeal2000,Neal2004,Bierkens2015}, and there is empirical evidence that these continuous-time MCMC algorithms are more efficient 
than their discrete-time counterparts
\cite[see e.g.][]{Bouchard:2016}. Similarly a continuous-time version of SMC has also been recently proposed \cite[]{Fearnhead/Latuszynski/Roberts/Sermaidis:2016}, which
involves propagating particles using a continuous-time Markov process. The original motivation for this was to be able to target distributions related to infinite-dimensional
stochastic processes, such as diffusions, without resorting to any time-discretisation approximations. However, we show below that one application of this methodology is to generate weighted-samples
from a target distribution of interest, giving an alternative interpretation of the recently proposed SCALE algorithm of \cite{Pollock/Fearnhead/Johansen/Roberts:2016}.

The purpose of this paper is to show that continuous-time MCMC and continuous-time SMC methods are linked through the fact that they both are based upon simulating
 continuous-time processes called piecewise-deterministic Markov processes. These are processes that evolve deterministically between a countable set of random
event times. The stochasticity in the process is due to the randomness in when these events occur, and possibly random dynamics at the event times. These processes
are natural building blocks of continuous-time Monte Carlo methods, as they involve a finite amount of computation to simulate as only a finite number of events and 
transitions are simulated in any fixed time-interval. 

Furthermore we aim to show that the methods that have been developed to date are just specific examples of a much wider class of potential continuous-time MCMC and SMC methods
that are based on piecewise-deterministic Markov processes. By giving an informal introduction to theory of piecewise-deterministic Markov processes, with emphasis on aspects
most relevant to the development of valid Monte Carlo methods, we hope to make the development of new continuous-time Monte Carlo methods more accessible and help stimulate
work in this area.

One aspect of continuous-time Monte Carlo that is particularly relevant for modern applications of Bayesian statistics is that they seem well-suited to big-data problems. If we
denote our target distribution by $\pi(\x)$ then the dynamics of these methods depend on the target through $\nabla \log \pi(\x)$. Now if $\pi(\x)$ is a posterior distribution,
then it will often be in product-form, where each factor relates to a data-point or set of data-points. 
This means that $\nabla \log \pi(\x)$ is a sum, and hence is easy to approximate unbiasedly using sub-samples of the data. It turns out we can use these unbiased
estimators within the continuous-time Monte Carlo methods without effecting their validity. That is, the algorithms will still target $\pi(\x)$. This is in comparison to
other discrete-time MCMC methods that use sub-sampling \cite[]{WellingTeh2011,Bardenet:2015,Ma:2015,Quiroz:2015}, where the approximation in the sub-sampling means that the algorithms will only target an approximation to $\pi(\x)$. It also
compares favourably with big-data methods that independently run MCMC on batches of data, and then combines the MCMC samples in some way \cite[]{Neiswanger:2013,Scott:2013,Srivastava:2015}. As the combination steps involved
will also introduce some approximation error.

The outline of the paper is as follows. In the next section we give an informal introduction Piecewise Deterministic Markov processes. Our aim is to cover key relevant concepts linked to
these processes whilst avoiding technical details. Those interested in a more rigorous introduction should see \cite{Davis:1984} and \cite{Davis1993}. Sections \ref{sec:CIS} and \ref{sec:CMCMC} then cover continuous-time
versions of SMC and MCMC respectively. These have been written so that either section could be read independently of the other. Our aim for each section is to introduce the continuous-time Monte
Carlo algorithm, show how it relates to a piecewise deterministic Markov process, and how we can use the theory for these processes to see that the Monte Carlo algorithms target the correct 
distribution. We also cover how these algorithms can be implemented using sub-sampling ideas, and highlight the importance of low-variance sub-sampling estimators for obtaining
highly efficient samplers for big-data. 

\section{Piecewise Deterministic Markov Processes}

The continuous-time versions of SMC, or sequential importance sampling, and MCMC that we will consider later are all examples of time-homogeneous piecewise-deterministic Markov processes. We will henceforth call these 
piecewise deterministic processes or PDPs. 

A PDP is a continuous-time stochastic process. Throughout we will denote the state of a PDP at time $t$ by $\Z_t$. 
The dynamics of the PDP involves random events, with deterministic dynamics between events and possibly random transitions at events. These dynamics are thus defined through specifying three quantities:
\begin{itemize}
\item[(i)] {\bf The deterministic dynamics.} We will assume that these are specified through an ordinary differential equation:
\begin{equation} \label{eq:ODE}
\frac{\mbox{d} z^{(i)}_t}{\mbox{d} t} = \Phi_i(\z_t),
\end{equation}
for $i=1,\ldots,d$, for some known vector-valued function $\Phi=(\Phi_1(\z),\ldots,\Phi_d(\z))$. This will lead to a deterministic transition function, 
so that the solution of the differential equation starting from value $\z_t$ and run for a time interval of length $s$ will give
\[
\z_{s+t}=\Psi(\z_t,s)
\] 
for some function $\Psi$. 
\item[(ii)] {\bf The event rate.} Events will occur singularly at a rate, $\lambda(\z_t)$, that depends only on the current position of the process. The probability of an event in interval in $[t,t+h]$ given
the state at time $t$, $\z_t$, is thus $\lambda(\z_t)h+o(h)$.
\item[(iii)] {\bf The transition distribution at events.} At each event the state of the process will change, according to some transition kernel. For an event at time $\tau$, if $\z_{\tau-}$ 
denotes  the state immediately prior to the event, then the state at time $\tau$ will be drawn from a distribution with density $q(\cdot|\z_{\tau-})$.  
\end{itemize}

To define a PDP process we will also need to specify its initial condition. We will assume that $\Z_0$ is drawn from some known distribution with density function $p_0(\cdot)$. 

\subsection{Simulating a PDP} \label{sec:Sim}

To be able to use a PDP as the basis of an importance sampling or MCMC algorithm, we will need to be able to simulate from it. A general approach to simulating a PDP is to iterate the following steps:
\begin{itemize}
\item[(S1)] Given the current time $t$ and state of the PDP $\z_t$, simulate the next event time, $\tau$ say.
\item[(S2)] Calculate the value of the process immediately before the next event time
\[
\z_{\tau-}=\Psi(\z_t,\tau-t).
\]
\item[(S3)] Simulate the new value of the process, immediately after the event, from $q(\z_\tau|\z_{\tau-})$.
\end{itemize}
The simulation algorithm is initiated with a current time $t=0$ and with $\Z_0$ drawn from the initial distribution of the process.
To simulate the process for a time interval $T$ these steps can be iterated until the first event time after $T$. If we wish to then simulate the value of the process at a time, $s$, between events we just find
the event time, $\tau$, immediately prior to $s$; the value of the process immediately after the event, $\z_{\tau}$; and then set
\[
\z_s=\Psi(\z_\tau,s-\tau).
\]
If $s$ is a time before the first event, then we would use $\tau=0$.

Below we will assume that our PDP has been chosen so that $\Psi(\cdot,\cdot)$ is known analytically and that the proposal distribution at events, $q(\cdot|\cdot)$ can be easily simulated from.
Thus the only challenging step to simulating a PDP will be simulating the next event in step S1. This involves simulating the next event in a time-inhomogeneous Poisson process. 

The first thing to note is that the event rate in (S1) can be written as a deterministic function of time, as the state dynamics are deterministic until the next event. 
If we are currently at time $t$ with state $\z_t$, then for any future time $t+s$ before the next event, the state will be $\z_{t+s}=\Psi(\z_t,s)$. Thus the event rate will be
\[
\lambda(\z_{t+s})=\lambda(\Psi(\z_t,s))=\tilde{\lambda}_{\z_t}(s),
\]
for a suitably defined function $\tilde{\lambda}_{\z_t}(\cdot)$. We can simulate the time until the next event, $s$, as the time of the first event in a Poisson process of rate $\tilde{\lambda}_{\z_t}(u)$. 

If the event rate is a simple function, then we can simulate events directly. If we define $\Lambda_{\z}(s)=\int_0^s \tilde{\lambda}_{\z}(u)\mbox{d}u$, then we can simulate such an event by (i) simulating $u$, the realisation of an exponential random with rate 1, and (ii) finding $s>0$ the solution
of $\Lambda_{\z}(s)=u$ \cite[e.g.][]{Cinlar:2013}. 

For more complicated rate functions either calculating $\Lambda_{\z}(s)$ or solving the equation in step (ii) may not be tractable. In such cases the most general approach to simulating event times is
 by thinning, or adaptive thinning \cite[e.g.][]{LewisShedler1979}. 

If we can upper-bound the event rate, $\tilde{\lambda}_{\z_t}(u)<\lambda^+(u)$, then thinning works by simulating possible events of a Poisson process of rate $\lambda^+(u)$ and 
accepting a possible event at time $u$ as an actual event with probability $\tilde{\lambda}_{\z_t}(u)/\lambda^+(u)$. The time of the first accepted event will be the time until the next event for our PDP.  
This requires the upper bound $\lambda^+(u)$ to be such that simulating events from a Poission process of rate $\lambda^+(u)$ is straightforward -- for example $\lambda^+(u)$ is constant or linear in $u$, or piecewise constant
or piecewise linear. Obviously the lower the bound $\lambda^+$, the more computationally efficient this approach will be.


\subsection{Analysing a PDP}

We now give informal introductions to some of the mathematical tools for analysing a PDP. 
These are introduced as they are used later to show that the continuous-time Monte Carlo methods we review have 
appropriate properties. For example, we introduce the generator of a PDP in the following section, and this is used to show that the continuous-time importance samplers of Section \ref{sec:CIS} 
produce properly weighted samples \cite[]{Liu/Chen:1998}. We then introduce the Fokker-Planck equation for a PDP, which can be used to showed that the continuous-time MCMC methods of Section \ref{sec:CMCMC} have the correct
invariant distribution. Understanding both the generator of a PDP and its Fokker-Planck equation is important if we wish to develop new versions of these continuous-time Monte Carlo methods.

\subsubsection{The Generator}

The generator of a continuous-time, time-homogeneous, Markov process is an operator that acts on functions of the state variable. We will denote the generator by $\mathcal{A}$. 
For suitable functions $f(\z)$, the generator is defined by
\[
\mathcal{A}f(\z) = \lim_{\delta t \rightarrow 0} \frac{\mbox{E}(f(\Z_{t+\delta t})|\Z_t=\z)-f(\z)}{\delta t}.
\]
The fact that the process is time-homogeneous means that the right-hand side does not depend on $t$. We can interpret the generator applied to a function $f(\z)$,
as giving the derivative of the expectation of $f(\Z_t)$ conditional on the current value of $\Z_t$. This quantity is only well-defined for certain functions, and the set of
such functions is called the domain of the generator. As the generator specifies how the expectation of any suitable function $f(\cdot)$ changes over time, it uniquely 
defines the dynamics of the underlying continuous-time stochastic process, in a similar way that knowing the moment generating function of a random variable will uniquely determine its distribution
\cite[]{EthierKurtz2005}.

If we are interested in the derivative of the expectation of a function of our PDP at a time $t$, then we can write this as
\[
\frac{\mbox{d} \mbox{E}(f(Z_t))}{\mbox{d} t} = \lim_{\delta t \rightarrow 0} \frac{\mbox{E}(f(\Z_{t+\delta t})-\mbox{E}(f(\Z_t))}{\delta t}= \lim_{\delta t \rightarrow 0} \mbox{E}_t\left(
\frac{\mbox{E}_{t+\delta t|t}(f(\Z_{t+\delta t})|\Z_t)-f(\Z_t)}{\delta t}\right),
\]
where in the last expression the inner expectation is with respect to $\Z_{t+\delta t}$ given $\Z_t$ and the outer expectation with respect to $\Z_t$. 
Assuming we can exchange the outer expectation and the limit we get
\begin{equation} \label{eq:deriv_exp}
\frac{\mbox{d} \mbox{E}(Z_t)}{\mbox{d} t}=\mbox{E}_t\left( \mathcal{A}f(\Z_t) \right).
\end{equation}
Thus the derivative of the expectation  of our function is the expectation of the generator applied to the function. 

\cite{Davis:1984} gives the generator for a piecewise deterministic process:
\begin{equation} \label{eq:gen_PDP}
\mathcal{A}f(\z) = \Phi(z) \cdot \nabla f(\z) + \lambda(\z) \int q(\z'|\z)[f(\z')-f(\z)] \mbox{d}\z',
\end{equation}
for functions $f(\cdot)$  such that $t \mapsto f(\Psi(z,t))$ is absolutely continuous.
The form of the generator has a simple interpretation. The first term on the right-hand side relates to the deterministic dynamics. For deterministic dynamics the generator is just the time-derivative of $f(\z_t)$, which by the product rule is
\[
\frac{\mbox{d}f(\z_t)}{\mbox{d}\z_t}=\sum_{i=1}^d \frac{\partial f(\z_t)}{\partial \z_t^{(i)} }\frac{\partial \z_t^{(i)}}{\partial t}=\Phi(\z) \cdot \nabla f(\z),
\]
where $\Phi(\z)$ is define in (\ref{eq:ODE}).
The second term on the right-hand side is the change in expectation at events. 
The probability of an event in time $[t,t+h]$ is $\lambda(\z_t)h+o(h)$ and the change in expectation conditional on event occuring, up to terms of $o(h)$, 
is given by the integral on the right-hand side.


\subsubsection{The Forward Operator and Fokker-Planck Equation}

We can define the adjoint of a generator of a continuous-time Markov process, $\mathcal{A}^*$, such that for suitable functions $g(\z)$ and $f(\z)$  
\[
\int g(\z) \mathcal{A}f(\z)\mbox{d}\z = \int f(\z) \mathcal{A}^*g(\z)\mbox{d}\z. 
\]
For function $g(\z)$ and $f(\z)$ which are both continuously differentiable and have compact support,
it is straightforward to show that the adjoint of the generator of a PDP (\ref{eq:gen_PDP}) is
\begin{equation} \label{eq:adj_PDP}
\mathcal{A}^* g(\z) = -\sum_{i=1}^d \frac{\partial \Phi_i(\z)g(\z)}{\partial \z^{(i)}} + \int g(\z') \lambda(\z')q(\z|\z')\mbox{d}\z' - g(\z)\lambda(\z)
\end{equation}
The first term equates to the adjoint of the first term of the generator, and is obtained using a straightforward integration by parts argument. To obtain this formula for
more general functions $f$ and $g$ requires more careful arguments to justify the change of order of integrations.

Now if we define the density function of our PDP at time $t$ to be $p_t(\z)$ then from (\ref{eq:deriv_exp}) we get that, for suitable function $f$, the derivate of the expectation of $f(Z_t)$ with respect to $t$ is
\[
\frac{\mbox{d} \mbox{E}(Z_t)}{\mbox{d} t}=\mbox{E}_t\left( \mathcal{A}f(\Z_t) \right)
=\int p_t(\z) \mathcal{A}f(\z)\mbox{d}\z = \int f(\z) \mathcal{A}^*p_t(\z)\mbox{d}\z.
\]
However we can equally write this derivative as
\[
\frac{\mbox{d} \mbox{E}(Z_t)}{\mbox{d} t}=\frac{\mbox{d}}{\mbox{d}t} \int p_t(\z) f(\z)\mbox{d}\z =\int \frac{\partial p_t(\z)}{\partial t}f(\z)\mbox{d}\z,
\]
assuming we can interchange differentiation and integration. This gives that
\[
\int \frac{\mbox{d} p_t(\z)}{\mbox{d}t}f(\z)\mbox{d}\z=\int f(\z) \mathcal{A}^*p_t(\z)\mbox{d}\z.
\]
As this holds for sufficiently many functions $f(\z)$ we get 
\[
\frac{\partial p_t(\z)}{\partial t} =\mathcal{A}^*p_t(\z).
\]
This is a partial differential equation for the distribution of the stochastic process, known as the Fokker-Planck {or Forward Kolmogorov} equation.
If $p(\z)$ is an invariant distribution of our PDP than it will satisfy $\mathcal{A}^*p(\z)=0$, which gives
\[
-\sum_{i=1}^d \frac{\partial \Phi_i(\z)p(\z)}{\partial \z^{(i)}} + \int p(\z') \lambda(\z')q(\z|\z')\mbox{d}\z' - p(\z)\lambda(\z) =0.
\]
The first term here relates to the change in probability mass caused by the deterministic dynamics, the second term relates to the probability flow into state $\z$ and the final term the probability flow out of state $\z$.
For an invariant distribution these will cancel for all states $\z$.


\section{Continuous-Time Sequential Importance Sampling} \label{sec:CIS}


A continuous-time version of sequential importance sampling was first developed to solve the problem of simulating from a diffusion. 
In this situation we have a diffusion process, $\X_t$, defined as the solution to an SDE
\[
\mbox{d}\X_t=\bb(\X_t)\mbox{d}t+\sigma(\X_t)\mbox{d}\B_t,
\]
where $\bb(\x)$ is the $d$-dimensional drift, $\B_t$ is $d$ dimensional Brownian motion, and $\sigma(\x)$ is a $d$ by $d$ matrix that defines the instantaneous variance of the process. 
We have an initial distribution $p_0(\x)$ for the diffusion, and wish to sample from the distribution of the process at some future time or times.
If we denote the density of this process at time $t$, by $p_t(\x)$, then the challenge is to sample from $p_t(\x)$ for diffusion processes where we cannot write down what $p_t(\x)$ is.

The Exact Algorithm \cite[]{Beskos/Roberts:2005,Beskos:2006} and its variants \cite[]{Beskos/Papaspiliopoulos/Roberts:2008,Pollock:2016} have given a number of algorithms for simulating from such a diffusion process, but only under strong conditions on the drift and instantaneous variance. For 
example it is commonly required that $\sigma(\x)$ is constant, and that the drift can be written as the gradient of some potential function. Almost all uni-variate diffusion processes can be transformed
to satisfy these requirements, but few multivariate diffusion processes can. 

Whilst we do not know what $p_t(\x)$ is for any $t>0$, we do know that it solves the Fokker-Planck equation for the diffusion
\[
\frac{\partial p_t(\x)}{\partial t}= -\sum_{i=1}^d \frac{\partial b_i(\x)p_t(\x)}{\partial x_i}+\frac{1}{2}\sum_{i=1}^d\sum_{j=1}^d \frac{\partial^2 \Sigma_{ij}(\x)p_t(\x)}{\partial x_i x_j},
\]
where $\Sigma=\sigma^T\sigma$. This motivates the following question: can we use our knowledge of the Fokker-Planck equation for the process
of interest in order to develop a valid importance sampling algorithm to sample from $p_t(\x)$?

The continuous-time importance sampling (CIS) procedure of \cite{Fearnhead/Latuszynski/Roberts/Sermaidis:2016}, which we describe below, will in fact enable us to do so. We will present it in a slightly more general form, in that we will use CIS to sample from a distribution 
$p_t(\x)$ that is the solution to a partial differential equation
\[
\frac{\partial p_t(\x)}{\partial t}=\mathcal{L}^{*}p_t(\x)
\]
for some known operator $\mathcal{L}^{*}$ and subject to a known initial condition $p_0(\x)$. This then allows for sampling from more general continuous-time Markov processes, 
where $\mathcal{L}^{*}$ would be the adjoint of the generator for that process. 

\subsection{The CIS Algorithm}

The idea of CIS is similar in spirit to a standard importance sampler. We will choose a tractable proposal process, for the problem of sampling from a diffusion this is most naturally chosen to be Brownian motion. 
This proposal process must have a know transition density which is simple to sample from. We will simulate paths from this proposal process up to time $t$, and then construct an importance sampling weight. 
The challenge is that we need to calculate an importance sampling weight without knowing $p_t(\x)$. The property we want from our importance sampler is that if we simulate a value and weight at time 
$t$,  $(\X, W)$, then for suitable functions $f(\x)$ we will have that $Wf(\X)$  will be an unbiased estimator of the expectation of $f(\X_t)$ for our target process,
\[
\mbox{E}(f(\X_t))=\int f(\x)p_t(\x)\mbox{d}\x.
\]

The original CIS algorithm was derived by taking a limit of a discrete-time sequential importance sampler. Below we give the general form of the resulting CIS process. The key to making this a valid importance
sampler is choosing the incremental weight (see step 2 below) appropriately. We will show how we can derive the form of the incremental weight by viewing the CIS process as a PDP, and using the generator
of the PDP to calculate expectations with respect to the CIS process.

\begin{figure}
 \centering
 \includegraphics[scale=0.7]{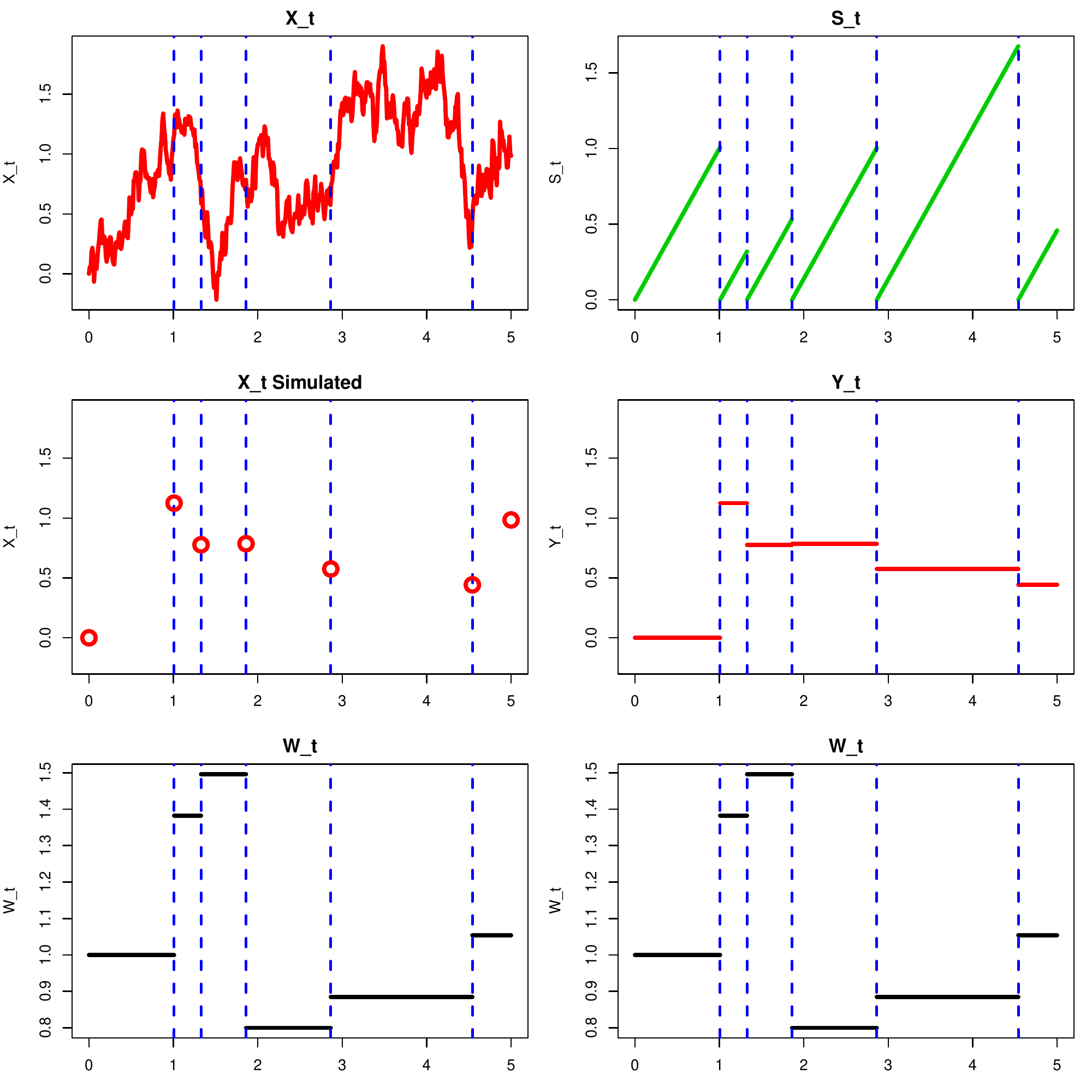}
 \caption{\label{Fig:CIS} Example of the CIS process. Left-hand column shows the original importance sampling process. Conceptually we think of a proposal process, $X_t$ (top plot), simulated in continuous-time. 
 However we need only simulate this process at event times (denoted by vertical dashed lines) and at times we are interested in. The middle plot shows these simulated values. 
 The weight process is shown in the bottom plot, and the weight changes only at event times. The right-hand column shows the corresponding components of our PDP process: $S_t$ (top); $Y_t$ (middle) and $W_t$ (bottom).}
\end{figure}

Denote the transition density of the proposal process over time interval $s$ as $q_s(x|y)$, and assume we have chosen an event-rate $\tilde{\lambda}(s)>0$ for $s>0$. The CIS algorithm is of the form:
\begin{itemize}
\item[(0)] Set $\tau=0$, $W_0=1$ and simulate $\X_0$ from the initial distribution of the target process, $p_0(\x)$.
\item[(1)]  Simulate a new event time $\tau'>\tau$, with the inter-event time $s=\tau'-\tau$ being drawn from a Poisson process with rate $\tilde{\lambda}(s)$. 
\item[(2)] If $\tau'>t$ then simulate $\X_t$ from $q_{t-\tau}(\cdot|\X_{\tau'})$, and set $W_t=W_{\tau'}$.
Otherwise simulate $\X_{\tau'}$ from $q_s(\cdot|\X_{\tau})$, update the weight
as
\begin{equation} \label{eq:weight}
W_{\tau'}=W_\tau \left[ 1+\frac{\rho(\X_{\tau'},\X_{\tau},s)}{\tilde{\lambda}(s)}\right],
\end{equation}
and set $\tau=\tau'$ and go to (1).
\end{itemize}
Step (0) is just an initialisation step. The idea of the algorithm is that we use random event times, simulated in step (1), at which we evaluate the proposal process. 
Based on the value of the process at both this event and the preceeding event we then update the importance sampling weight. As is standard in sequential importance sampling, 
the new weight (calculated in step 2) is just the old weight multiplied by an incremental weight. Figure \ref{Fig:CIS} gives an example of the output of this algorithm.

The key to making this a valid importance sampling algorithm is working out an appropriate 
form of the incremental weight. Without loss of generality we have written the incremental weight in the form given inside the square brackets in (\ref{eq:weight}), as this will simplify the derivation later. 
To specify the incremental weight, we need to appropriately choose the
function $\rho$.

\subsubsection{CIS as a PDP}

It can be seen that the CIS process is just a piecewise deterministic Markov process. The only randomness in the process is the timing of the events and the transitions at the events. 
We can formalise this by defining a PDP with state $\Z_t=(\Y_t,W_t,S_t)$ where $\Y_t$ is the value of the CIS process simulated at the most recent event prior to $t$, $W_t$ is the importance sampling weight 
at time $t$ and $S_t$ is the time since the most recent event. See Figure \ref{Fig:CIS} for an example.

This PDP has deterministic dynamics given by differential equations
\[
\frac{\mbox{d} \Y_t}{\mbox{d}t}=0,~~~~~~
\frac{\mbox{d} W_t}{\mbox{d}t}=0,~~\mbox{ and }~~
\frac{\mbox{d} S_t}{\mbox{d}t}=1.
\]
Events occur at a rate $\lambda(\Z_t)=\tilde{\lambda}(S_t)$. At an event at time $\tau$ we simulate $\Y_{\tau}$ from $q_{S_{\tau-}}(\cdot|\Y_{\tau-})$, set $S_\tau=0$ and update $W_{\tau}$ as described in (\ref{eq:weight}).

The generator of $\Z_t$, given by (\ref{eq:gen_PDP}), is
\begin{equation} 
\label{eq:generator-CIS} \mathcal A h(\y, w, s) = \frac{\partial h(\y, w, s)}{\partial s} + \tilde \lambda(s) \int \left\{ h \left(\y', w \left( 1 + \frac{\rho(\y', \y, s)}{\tilde \lambda(s)} \right), 0 \right) - h(\y, w, s) \right\} q_{s}(\y' \mid \y) \ d \y'.
\end{equation}
Given the state at time $t$, $\Z_t=(\Y_t, W_t, S_t)$, 
our estimate of the expectation of $f(\X_t)$ under our target process will be $W_t f(\X)$, where $\X$ is drawn from $q_{S_t}(\cdot \mid \Y_t)$. 
The requirement that our algorithm is a valid importance sampler then becomes that
\[
\mbox{E}_{\X,\Z}\left( W_tf(\X) 
\right):=\mbox{E}_{\Z}\left( W_t\int f(\x)q_{S_t}(\x|\Y_t) \mbox{d}\x
\right)= \int f(\x)p_t(\x)\mbox{d}t.
\]
Here the first expectation is with respect to $\Z$, the PDP, and $\X$ the simulated value for $\X_t$, and the second expectation is just with respect to $\Z$. 

As this must hold for all appropriate functions $f$, we need that for almost all $\x$
\[
\mbox{E}\left( W_t q_{S_t}(\x|\Y_t)\right) =p_t(\x).
\]
We will denote the left-hand side by $\tilde{p}_t(\x)$. Due to the initialisation of the CIS algorithm we have $\tilde{p}_0(\x)=p_0(\x)$. 
Thus we need to choose the incremental weights such that $\tilde{p}_t(\x)$ satisfies the Fokker-Planck equation for the target  process
\begin{equation} \label{eq:FP}
\frac{\partial \tilde{p}_t(\x)}{\partial t}=\mathcal{L}^{*} \tilde{p}_t(\x).
\end{equation}

\subsubsection{Obtaining the Incremental Weight}
We will give an informal outline of how to derive the function $\rho$. Throughout this argument we will assume that we can interchange expectation with various operators. See 
\cite{Fearnhead/Latuszynski/Roberts/Sermaidis:2016}, and the discussion below, for conditions under which this is valid.

We need to choose $\rho$ so that (\ref{eq:FP}) holds. The right-hand side of (\ref{eq:FP}) is just
\[
\mathcal{L}^{*} \tilde{p}_t(\x)=\mbox{E}\left( \mathcal{L}^{*} W_tq_{S_t}(x|\Y_t) \right).
\]
The left hand side of (\ref{eq:FP}) is the derivative of an expectation, and thus can be written in terms of the generator of the PDP, $\mathcal{A}$,
\[
\frac{\partial \tilde{p}_t(\x)}{\partial t}=\frac{\partial \mbox{E}(W_tq_{S_t}(x|\Y_t))}{\partial t}
= \mbox{E}\left( \mathcal{A} W_tq_{S_t}(x|\Y_t)) \right). 
\]
So both the left and right hand sides of (\ref{eq:FP})  can be written as an expectation of a function of the PDP. For these expectations to be equal it is sufficient for the two functions to be equal to each other.

Using the form of the generator for the CIS process (\ref{eq:generator-CIS}) we get that for current state $\z=(\y,w,s)$,
\begin{eqnarray*}
\mathcal{A} w q_{s}(\x|\y) 
&=& \frac{\partial wq_{s}(\x|\y)}{\partial s} + \tilde{\lambda}(s) \int \left\{w\left( 1 + \frac{\rho(\y', \y, s)}{\tilde \lambda(s)} \right)q_0(\x|\y')-q_{s}(\x|\y) \right\} q_{s}(\y' \mid \y)\mbox{d} \y' \\
&=& w \frac{\partial q_{s}(\x|\y)}{\partial s}+ w\tilde{\lambda}(s) \left\{  \int  \left( 1 + \frac{\rho(\y', \y, s)}{\tilde \lambda(s)} \right) q_0(\x|\y')    q_{s}(\y' \mid \y)  \mbox{d} \y' -q_{s}(\x|\y)\int q_{s}(\y' \mid \y)\mbox{d} \y'             \right\}\\
&=&w \frac{\partial q_{s}(\x|\y)}{\partial s}+  \tilde{\lambda}(s)\left[ w q_s(\x|\y)\left(1+\frac{\rho(\x;\y,s)}{\tilde{\lambda}(s)}\right) - w q_s(\x|\y)\right] \\
&=&w\frac{\partial  q_s(\x|\y)}{\partial  s}+ w q_s(\x|\y)\rho(\x,\y,s).
\end{eqnarray*}
The above argument is informal, as for $s=0$ (a state visited every time a jump occurs) the function
$q_s(x|y)$ does not exist. However it acts, informally, like a dirac delta function.
So for any function $h(\y')$ we have $\int q_0(\x|\y')h(\y') \mbox{d}\y'=h(\x)$, and this is used above.
We refer the reader to
\cite{Fearnhead/Latuszynski/Roberts/Sermaidis:2016} for formal justification
of this step and others.

Thus for the two functions of the PDP that we are taking expectations of to be equal we need
\[
w\frac{\partial  q_s(\x|\y)}{\partial  s}+ w q_s(\x|\y)\rho(\x,\y,s)=\mathcal{L}^{*} wq_{s}(\x|\y),
\]
which can be re-arranged to give
\begin{equation} \label{eq:rho}
\rho(\x,\y,s)=\frac{ \mathcal{L}^{*} q_{s}(\x|\y)-\frac{\partial  q_s(\x|\y)}{\partial  s}}{ q_s(\x|\y)}.
\end{equation}
The incremental weight is then $1+\rho(\x,\y,s)/\tilde{\lambda}(s)$.

The form of this incremental weight is quite intuitive. It is based on the difference between how transition densities change under the target process and under the proposal process. The optimal proposal process would, 
obviously, be the target process. For this choice $\partial q_s(\x|\y)/\partial s=\mathcal{L}^{*} q_{s}(\x|\y)$, $\rho(\x,\y,s)=0$, and the importance sampling weights would always be equal to 1. As expected,
a proposal process that more closely mimics the target process will have less variable incremental weights.
Furthermore we see a trade-off in the choice of the event rate $\tilde{\lambda}(s)$, as larger values of this rate will lead to more events and a higher computational cost,
but reduce the variance in the incremental weight. If we were to double the event rate, we would double the expected number of events, but the variance of the incremental weight at an event would reduce by a factor of 4. 
We expect that the net effect is an overall reduction of Monte Carlo variance in the weights but an increase in computational cost.

One issue with the CIS algorithm is that, for certain combinations of target and proposal processes, it can be possible to get negative weights
\cite[similar, for example, to the Russian roulette sampler of][]{lyne2015}. This can occur if it is possible for $\rho(\x,\y,s)$ to be smaller
than $-\tilde{\lambda}(s)$. For the applications of CIS to simulating from a general diffusion process that are described in \cite{Fearnhead/Latuszynski/Roberts/Sermaidis:2016}, $\rho(\x,\y,s)$ cannot be bounded below, and thus
negative weights are possible. Obviously the probability of getting negative weights can be controlled, with larger $\tilde{\lambda}(s)$ values reducing this probability. This in turn leads to important theoretical and practical 
issues.  The above argument for deriving the incremental weight required the interchange of expectation and operators. The main condition for justifying this is that the we need the importance sampling weights to be such 
that $\E(|W_t|)$ is finite for any $t>0$. This condition may not hold due to the possibility of negative weights. In fact, \cite{Fearnhead/Latuszynski/Roberts/Sermaidis:2016}, show that some naive implementations
of CIS will not have importance sampling weights that satisfy this condition. Furthermore, for problems where the target process is a diffusion, they give sufficient conditions on both the proposal process and the event
rate that will ensure $\E(|W_t|)$ is finite. In general, and within the more specific context of diffusion proposals and targets, \cite{Fearnhead/Latuszynski/Roberts/Sermaidis:2016} provide conditions under which 
$W_t$ has $p$-th moments for $p\ge 1$. The paper also gives practically implementable and intuitive strategies for ensuring that these moments exist.

\subsection{Continuous-time Sequential Monte Carlo}

To date we have reviewed a continuous-time importance sampling algorithm. This is most naturally viewed as an extension of sequential importance sampling \cite[]{Liu/Chen:1998} to continuous-time. However it is then possible 
to extend this to a continuous-time version of SMC. All we need to do is to jointly simulate multiple CIS processes, and then introduce resampling steps. The simplest implementation of
this is to choose a set of resampling times, say $h,2h,3h,\ldots,Kh$, and a number of ``particles'', $N$. The continuous-time SMC algorithm will then proceed as follows:
\begin{itemize}
 \item[(0)] {\bf Initiate}. Simulate $\x_0^{(1)},\ldots,\x_0^{(N)}$ independently from $p_0(\x)$. Set $w_0^{(i)}=1$ for $i=1,\ldots,N$. Set $t=0$
 \item[(1)] {\bf Propagate and Reweight}. For $i=1,\ldots,N$ run the CIS process for a time interval of length $h$, with initial values $\x_t^{(i)}$ and $w_t^{(i)}$. Denote
 the output of the CIS process at time $u=t+h$ by $\tilde{\x}_u^{(i)}$ and $\tilde{w}_u^{(i)}$.
 \item[(2)] {\bf Resample}. For $i=1,\ldots,N$ sample $k_i$ from $1,\ldots,N$ with probabilities proportional to $|\tilde{w}_u^{(1)}|,\ldots,|\tilde{w}_u^{(N)}|$. Set $\x_u^{(i)}=\tilde{\x}_u^{(k_i)}$ and 
 \[
 w_u^{(i)}=\frac{\tilde{w}_u^{k_i}}{|\tilde{w}_u^{k_i}|} \frac{1}{N}\sum_{j=1}^N |\tilde{w}_u^{(j)}|.
 \]
 \item[(3)] {\bf Iterate}. Set $t=u$. If $t<K\tau$ go to step (1).
\end{itemize}
The resampling step is different from standard resampling used in SMC to allow for the possibility of negative weights. The form of the weight after resampling is defined so that resampling is unbiased, and this requires
the sign of the weight assigned to any particle value to be unchanged. It is simple to extend the above use of resampling to allow for lower-variance resampling schemes 
\cite[]{Kitagawa:1996,Carpenter/Clifford/Fearnhead:1999,Doucet/Godsill/Andrieu:2000} in step (2), and to allow resampling
times to depend on the SMC output, for example to be times when the effective sample size of the weights drops below some threshold \cite[]{Liu/Chen:1995}. 

The beneficial effect of resampling will be less when we have negative weights. For example, it is easy to show that $\E(|W_t|)$ is unchanged by resampling. Thus resampling cannot counteract any increase in $ \E(|W_t|)$ 
with $t$, and this increase will necessarily imply deterioration of Monte Carlo performance with $t$. Thus the good long-term stability properties that SMC with resampling often has
\cite[]{DelMoral:2001,Douc:2014} will not be possible in the
presence of negative weights. This issue with negative weights is well-known within related quantum Monte Carlo methods \cite[]{Foulkes:2001}, where it is termed the fermion-sign problem. 

\cite{Fearnhead/Latuszynski/Roberts/Sermaidis:2016} suggest alternative resampling approaches for step (2) that can reduce $\E(|W_t|)$ at resampling times, and thus may lead to 
long-term stability. Alternatively, we need to choose event rates in CIS to be sufficiently large that negative weights are rare over the time-scales that we wish to run an SMC algorithm for.





\subsection{CIS for Big Data: the SCALE Algorithm} \label{S:SCALE}

Recently \cite{Pollock/Fearnhead/Johansen/Roberts:2016} presented SCALE, an algorithm for sampling from a posterior distribution. The original derivation of SCALE
was based on constructing a killed Brownian-motion process whose quasi-stationary distribution is the posterior distribution. The SCALE algorithm then samples from this quasi-stationary distribution. 
A key property of SCALE is that it only needs to use a small sub-sample of the data at each iteration of the algorithm, and thus it is suitable for large data applications.

Whilst the original derivation is very different, we show here that SCALE can be viewed as a CIS, or continuous-time SMC, algorithm. 
Our setting is that we wish to sample from a posterior distribution which we will assume can be written as
\[
 \pi(\x)\propto \prod_{i=1}^n \pi_i(\x),
\]
where, to keep notation consistent with our presentation of CIS, $\x$ is the parameter vector. Here $\pi_i(\x)$ is the likelihood for the $i$th observation multiplied by the $1/n$th power of the prior. 
As is common to Bayesian inference, the posterior distribution is only known up to a constant of proportionality. We 
wish to develop a Monte Carlo algorithm for sampling from this posterior that has good computational properties for large $n$. 

The idea of SCALE and its link to CIS is as follows. We will use CIS to target a stochastic process that has $\pi(\x)$ as its stationary distribution. To implement CIS we only need to know the Fokker-Planck equation
for this process. If we run CIS (or in practice a continuous-time SMC version) then after a suitable burn-in period this will gives us weighted samples from $\pi(\x)$. 
A key property of the CIS algorithm is that the incremental weights depend on the posterior only through $\log \pi(\x)$. This is a sum, and it is easy to use sub-sampling to unbiasedly estimate this sum.
Thus to deal with potentially large data we will  implement a random weight version of CIS \cite[]{Fearnhead/Papaspiliopoulos/Roberts:2008}, where we use sub-sampling to estimate the incremental weights. Using
unbiased random weights leads to a valid importance sampler, but one with an increased Monte Carlo error. The next ingredient to the SCALE algorithm is to use control variates to reduce the variance of our
sub-sampled estimates of the incremental weights, which in turn helps to control the overall Monte Carlo error of the algorithm. Finally \cite{Pollock/Fearnhead/Johansen/Roberts:2016} use ideas from the Exact Algorithm to avoid the possibility of negative
weights. We now detail each of these steps.

The first step to SCALE is the choice of a stochastic process that has $\pi(\x)$ as its asymptotical distribution. We need specify this process through its Fokker-Planck equation. SCALE uses the stochastic process
for which
\[
 \frac{\partial p_t(\x)}{\partial t}=\frac{1}{2}\sum_{i=1}^d \frac{\partial^2p_t(\x)}{\partial x_i^2} - \frac{1}{2 \pi(\x)}\left(\sum_{i=1}^d\frac{\partial^2 \pi(\x)}{\partial x_i^2}\right)  p_t(\x).
\]
It is trivial to see that $\pi(\x)$ is an invariant distribution for this stochastic process, as on substituting $p_t(\x)=\pi(x)$ the two terms on the right-hand side cancel. The actual underlying stochastic process can be
interpreted as Brownian motion with killing, conditioned on survival: see \cite{Pollock/Fearnhead/Johansen/Roberts:2016} for more details. 

If we implement CIS for this target process, and use Brownian motion as the proposal distribution, we have
\[
 q_s(\x|\y)=\left(\frac{1}{\sqrt{2\pi s}}\right)^d \exp\left\{- \sum_{i=1}^d \frac{(x_i-y_i)^2}{2s} \right\},
\]
and
\[
 \frac{\partial q_s(\x|\y)}{\partial s}=\frac{1}{2}\sum_{i=1}^d \frac{\partial^2 q_s(\x|\y)}{\partial x_i^2}.
\]

Thus from (\ref{eq:rho}), the function that determines the incremental weights is 
\[
 \rho(\x,\y,s)= - \frac{1}{2 \pi(\x)}\sum_{i=1}^d \frac{\partial^2 \pi(\x)}{\partial x_i^2} = - \frac{1}{2}\sum_{i=1}^d \left[ \frac{\partial^2 \log \pi(\x)}{\partial x_i^2}+ 
  \left(\frac{\partial \log \pi(\x)}{\partial x_i} \right)^2 \right].
\]
The right-hand side depends on $\pi(\x)$ only through derivatives of
\[
 \log \pi(\x) = \mbox{constant}+\sum_{i=1}^n \log\pi_i(\x).
\]
Importantly these derivatives do not depend on the unknown normalising constant of $\pi(\x)$. Furthermore as they are sums, it is simple to unbiasedly estimate the derivatives. For example
given $j$ and $k$, two independent draws from an uniform distribution on ${1,\ldots,n}$, we can estimate $\rho(\x,\y,s)$ by
\begin{equation} \label{eq:SS1}
 -\frac{1}{2}\sum_{i=1}^d \left[ n \frac{\partial^2 \log \pi_j(\x)}{\partial x_i^2}+ 
  n^2\left(\frac{\partial \log \pi_j(\x)}{\partial x_i} \right)\left(\frac{\partial \log \pi_k(\x)}{\partial x_i} \right)\right].
\end{equation}

We can reduce the variance of our estimate of $\rho$, and hence of the incremental weights, using control variates. \cite{Pollock/Fearnhead/Johansen/Roberts:2016} suggest using a pre-processing step
that finds a values $\hat{\x}$ close to the posterior mode \cite[for similar ideas, see][]{Bardenet:2015}. We precalculate the first and second derivates of $\log \pi(\x)$ at $\hat{\x}$, and calculate the value, $\hat{\rho}$, of $\rho$ at $\hat{\x}$. 
We then estimate of $\rho$ at $\x$ as 
\begin{equation} \label{eq:rho_cv}
-\frac{n}{2}\sum_{i=1}^d\left\{\left[\frac{\partial^2 \log \pi_j(\x)}{\partial x_i^2}- \frac{\partial^2 \log \pi_j(\hat{\x})}{\partial x_i^2}\right]
 +n\left[ \frac{\partial \log \pi_j(\x)}{\partial x_i} - \frac{\partial \log \pi_j(\hat{\x})}{\partial x_i}\right]
 \left[  \frac{\partial \log \pi_j(\x)}{\partial x_i} - \frac{\partial \log \pi_j(\hat{\x})}{\partial x_i} + 2\frac{1}{n}\frac{\partial \log \pi(\hat{\x})}{\partial x_i}\right]\right\}
 +\hat{\rho}
\end{equation}
where, as before, $j$ and $k$ are independent draws from an uniform distribution on ${1,\ldots,n}$. The idea behind this control-variate approach is that if $\hat{\x}$ is within $1/\sqrt{n}$ of the posterior
mode, then with high-probability, at stationarity the CIS process will be at an $\x$ value that is within $1/\sqrt{n}$ of $\hat{\x}$. If the first and second derivatives of the $\log \pi_j$s are well-behaved this means
that the terms in the square brackets will be $O_p(1/\sqrt{n})$, and thus $\rho$ will be $O(n)$ with high probability. Thus compares well with the naive subsampling estimator of $\rho$ (\ref{eq:SS1}), which will be 
$O_p(n^2)$.

Finally, to avoid negative weights, the SCALE algorithm then uses ideas from \cite{Beskos/Papaspiliopoulos/Roberts:2008} and \cite{MaCiS:BJ08}, to simulate the proposal process in such a way that we know an upper and lower bound the process takes within a given time-interval.
With such bounds we can then choose our event-rate $\lambda$ sufficiently high that negative weights do not occur. This approach does come with a computational cost, as simulating Brownian motion together
with such a bound can be an order of magnitude, or more, slower than just simulating Brownian motion. Below we investigate the feasibility of implementing a version of SCALE that allows for negative weights, but 
that chooses event-rates to be sufficiently large that they are rare. 

\subsection{Extensions}

Here we give two examples of how we can use the theory for PDPs to easily obtain generalisations of the basic CIS and SCALE algorithms. 

\subsubsection{Alternative Proposal Distributions}

With any importance sampling approach, the choice of proposal distribution can have a substantial impact on the resulting Monte Carlo properties. The original derivation of CIS used the idea of a continuous-time
stochastic process that was being used as a proposal process.
However our derivation of the CIS algorithm has not actually required us to specify a proposal process, just a suitable family of transition densities, $q_s(\x|\y)$. This family needs to have certain properties, 
such as being differentiable with respect to $s$, so that the incremental weight (\ref{eq:rho}) can be calculated. This appears to make it easy to consider alternative proposals, as we only need to specify an appropriate family of densities. For example, in standard importance sampling
applications it is often recommended to use heavy-tailed proposals, so for CIS it is natural to consider transition densities that are $t$-distributed, such as
\begin{equation} \label{eq:tprop}
 q_s(\x|\y) \propto s^{-d/2} \frac{1}{(1+\frac{1}{\nu}\sum_{i=1}^d(x_i-y_i)^2/s)^{(\nu+d)/2}},
\end{equation}
for some appropriately chosen degrees of freedom $\nu>0$.

\subsubsection{Alternative SCALE Processes}

We can also develop alternatives to the SCALE algorithm that differ in terms of the underlying stochastic process that they target. For example we can target the Langevin diffusion, for which the 
Fokker-Planck equation is
\[
 \frac{\partial p_t(\x)}{\partial t} = \frac{1}{2}\sum_{i=1}^d \frac{\partial^2p_t(\x)}{\partial x_i^2}-\frac{1}{2}\sum_{i=1}^d\left[\frac{\partial}{\partial x_i}\left(p_t(\x)\frac{\partial \log \pi(\x)}{\partial x_i} \right) \right].
\]
Again, it is simple to see, by substituting $p_t(\x)=\pi(\x)$ on the right-hand side, that $\pi(\x)$ is the invariant distribution for this process. If we implement CIS to sample from this Langevin diffusion,
and use Brownian Motion as the proposal distribution, we get that the incremental weights are $1+\rho(\x,\y,s)/\tilde{\lambda}(s)$ with
\[
 \rho(\x,\y,s)=-\frac{1}{2}\sum_{i=1}^d\left[ 
 \frac{(y_i-x_i)}{s}\frac{\partial \log \pi(\x)}{\partial x_i}
 + \frac{\partial^2 \log \pi(\x)}{\partial x_i^2}.
 \right]
\]
We can unbiasedly estimate this using a sub-sample of size 1. For example, if $j$ is drawn uniformly from $\{1,\ldots,n\}$, one unbiased estimator is
\[
 \hat{\rho}(\x,\y,s)=-\frac{n}{2}\sum_{i=1}^d\left[ 
 \frac{(y_i-x_i)}{s}\frac{\partial \log \pi_j(\x)}{\partial x_i}
 + \frac{\partial^2 \log \pi_j(\x)}{\partial x_i^2}
 \right].
\]
Alternatively we can develop lower-variance esimators using control variates.
The incremental weight is easier to estimate than the incremental weight in SCALE as it does not involve a square of the gradient of the log-posterior. However, the $(y_i-x_i)/s$ term has a variance, under the
Brownian motion proposal, that is $1/s$, and care is needed to control this term for small inter-event times, $s$. 

\subsection{Example: Inference for Mixture Models} \label{sec:MM1}

To give an indication of how the SCALE algorithm of Section \ref{S:SCALE} works, some of the issues with its implementation, and the importance of using control-variates with the sub-sampling estimator or $\rho$, we
will consider a simple example. Assume we have IID data from a mixture distribution, where for $j=1\ldots,n$
\[
 Y_j \sim \left\{ \begin{array}{cl} \mbox{N}(0,10^2) & \mbox{with probability $p$}, \\
 \mbox{N}(x,1^2) & \mbox{otherwise.}
 \end{array}\right.
\]
Our interest is inference for $x$, and we assume a Gaussian prior with mean 0 and variance 4. In the following we simulate data with $x=4$.

\begin{figure}
 \centering
 \includegraphics[scale=0.8]{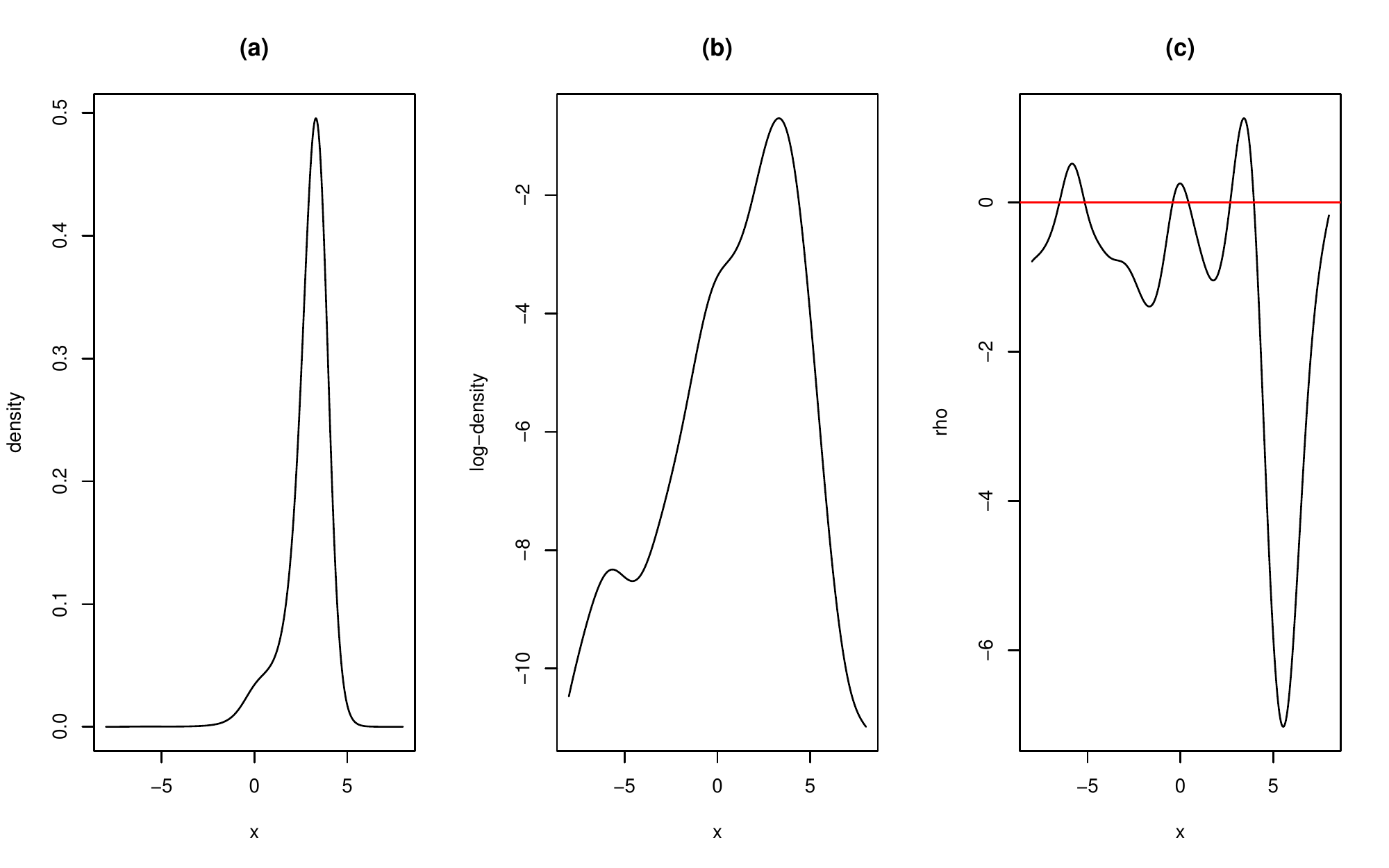}
 \caption{\label{Fig:1} The posterior distribution (a); the log-posterior (b); and $\rho$ (c) for a data set of size 150 from the mixture model.}
\end{figure}

Initially we implemented the SCALE algorithm without any subsampling on a data-set of size $n=150$ and $p=0.95$. Figure \ref{Fig:1} show the posterior distribution for $x$ for this data set, the log-posterior and how $\rho$ varies
as a function of $x$. Note that the average value of $\rho$ at stationarity is 0, so values where $\rho$ is greater than or less than 0 show regions where weights of particles will increase or decrease relative to
the average weight. The key point in these figures is that while both the posterior and log-posterior are relatively well-behaved, $\rho$ has multiple pronounced modes. 

This multi-modality can cause issues with mixing.
For example a particle that is currently at a value of $-5$ will have to traverse a prolonged region where $\rho$ is negative to reach the main region of posterior mass at values of $x$ close to 4. As it traverses
this region its weight will reduce substantially, and it is likely to be lost due to resampling. Thus we could have relatively rare movement of particles from one mode of $\rho$ to another.

\begin{figure}
 \centering
 \includegraphics[scale=0.8]{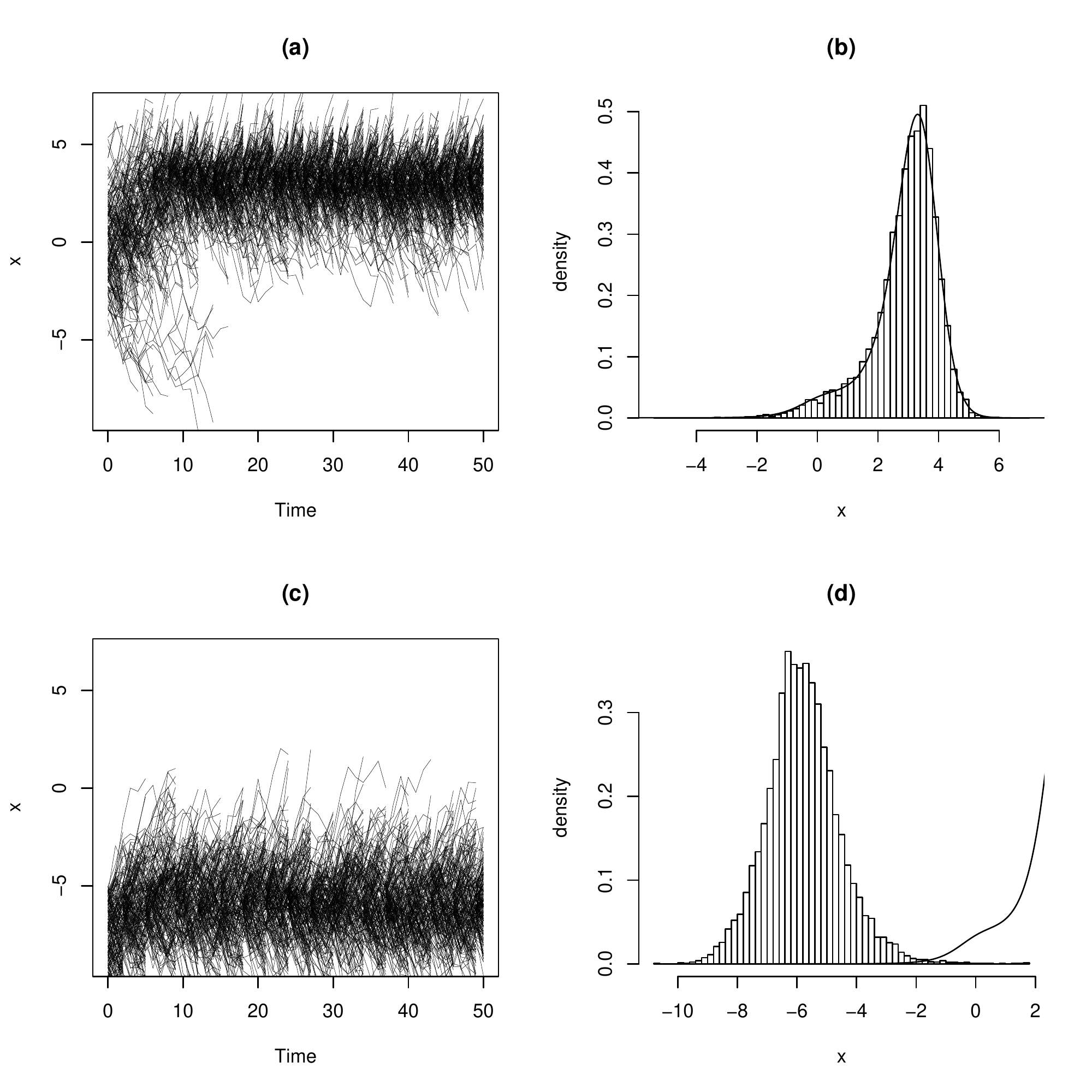}
 \caption{\label{Fig:2} Evolution of particles up to time 50 (left-hand column) and estimate of posterior (right-hand column) from SCALE algorithm. Top row is for particles initiated from prior and bottom row for particles
 initiated uniformly on $[-10,-5]$. Estimates of posterior are show as histograms based on the weighted particles from time 25 to 100, and the true posterior is overlain.}
\end{figure}

In practice this means that initialisation of the SCALE algorithm can be crucial. We implemented SCALE with 200 particles over a time interval of length 100. We considered resampling at every integer time-point, and 
resampled if the effective sample size of the weights was less than 100. We ran SCALE with a constant event rate of 12, and observed no negative weights. The performance of SCALE appeared quite sensitive to the event rate,
with rates of 10 and less giving noticeably worse performance. Figure \ref{Fig:2} shows output from two runs of SCALE, one 
initialised with particles drawn from the prior, and the other initialised with particles drawn uniformly on $[-10,-5]$. Figure \ref{Fig:2} (a) and (c) show the evolution of the particles over the first 50 time-steps. 
They evolve according to Brownian motion, but accrue or lose weight depending whether $\rho$ is positive or negative at the particle values at event times. Particles with low weights tend to be lost at resampling times. 
For the case where we initialise particles from the prior, we see particles are quickly lost in regions aroung $x=0$ where $\rho$ is negative. 
More slowly they are also lost from regions around $x=-5$ which is a local-maxima in $\rho$, as these particles, on average, attain a smaller weight than those particles close to the posterior mode. By about time 15 
the SCALE algorithm appears to have converged, and Figure \ref{Fig:2} (b) shows that it gives a good approximation to the true posterior distribution. By comparison, when SCALE is initialised with no particles close
to the posterior mode, the particles appear to get stuck close to local mode of $\rho$ at $x=-5$. 

\begin{figure}
 \centering
 \includegraphics[scale=0.6]{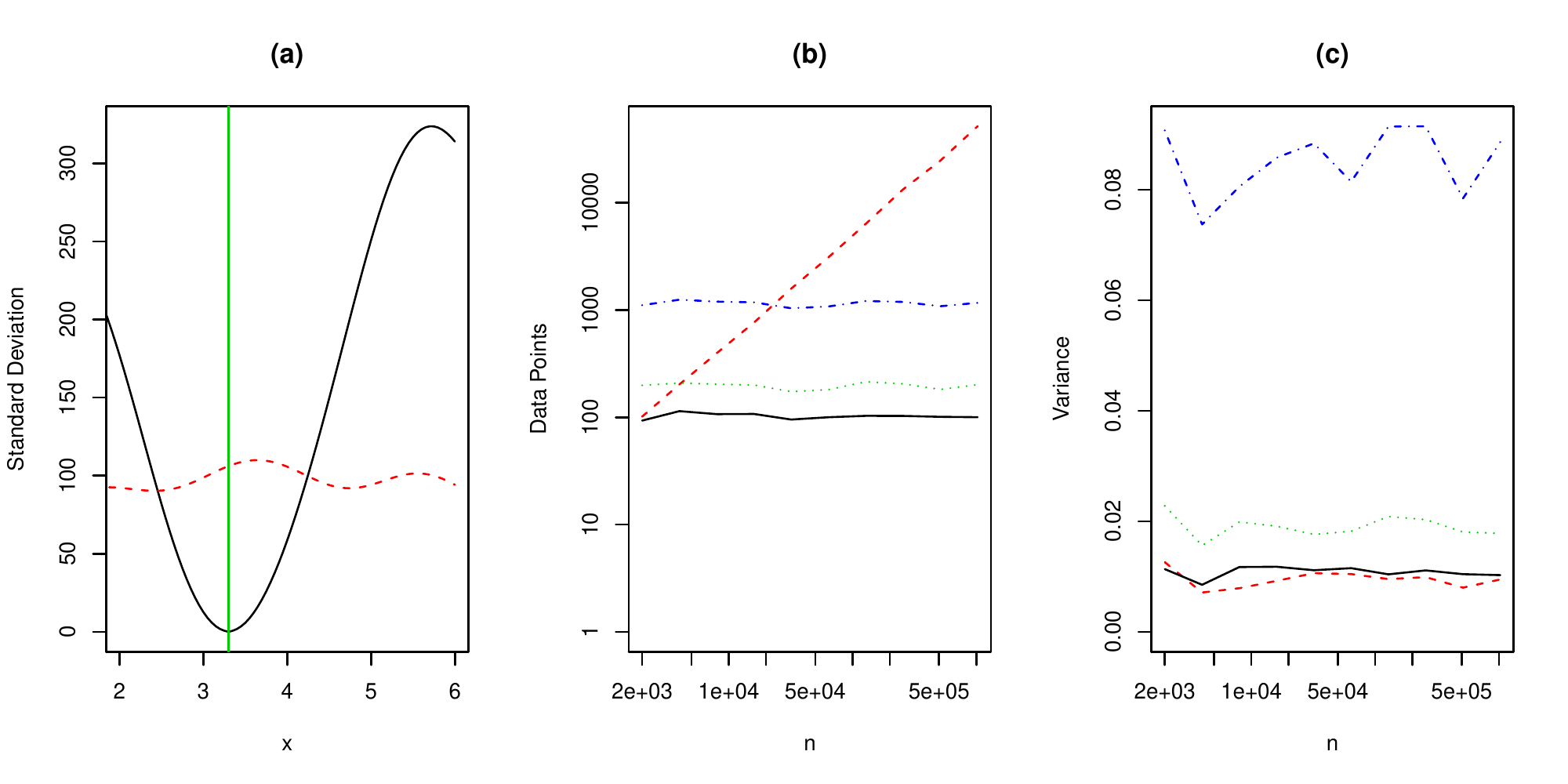}
 \caption{\label{Fig:3} Variance of estimate of $\rho$ using subsampling with (black full-line) and without (red dashed line) control variates (plot a); the vertical line shows the value of $\hat{x}$. Computational
 cost of estimating $W_h$, measured in terms of number of data point accessed, as a function of $n$ (plot b), and variance of $W_h$ as a function of $n$ (plot c). For these latter two plots, results are shown for  
 no subsampling (red dashed line) and subsampling with $\hat{x}$ as the posterior mode (black full-line) the mode plus posterior standard deviation (green dotted line) and mode plus three times the standard deviation 
 (blue dot-dash line). Results calculated from $2,000$ estimates of $W_h$ for each method and each value of $n$.}
\end{figure}

We now turn our attention to the possibly benefits of using subsampling to estimate $\rho$. Firstly to show the difference between subsampling with and without control variates, Figure \ref{Fig:3}(a) shows the variance 
of our the sub-sampling estimate of $\rho$ for our data set of size $150$. The control variate estimator (\ref{eq:rho_cv}) was implemented with $\hat{x}$ set to be the posterior mode. We see a substantially 
lower variance for $x$
close to $\hat{x}$ when we use control variates. Though as $x$ moves away from $\hat{x}$ the variance increases, and eventually is worse than not using control-variates. The key to why control variates works for
large data is that if $\hat{x}$ is close to the posterior mode then $x$ will be very close to $\hat{x}$ with high-probabililty. In such cases the proportionate reduction in variance will be $O(n)$ --
and thus the gains of using control variates will increase for larger data sets.

To gain insight into the benefit of using sub-sampling we looked at the variance of the weight at time $h$, $W_h$, of a
particle sampled from the true posterior, for different values of $n$ both with and without subsampling. For a fixed variance of $W_h$ we wanted to see how many data-points need to be processed by each algorithm.
So without sub-sampling, each event requires access to all $n$ data points, whereas with sub-sampling each event requires us to access only 2 data points. We found that sub-sampling without control variates performed
substantially worse than the two alternatives, with increasingly poor performance as $n$ increases. Thus we focus on comparing no subsampling and subsampling with control variates.

As we increase $n$ the posterior standard deviation will decrease at a rate $1/\sqrt{n}$, and thus we choose $h=1/n$ so that the distance moved by
the particle will also be of order $1/\sqrt{n}$. When using no-subsampling we chose the event rate to be $n/2$. For sub-sampling with control variates we set $\hat{x}$ to be the posterior mode and let 
the event rate depend on the value of the particle at the most recent event, $x'$ say, and be of the form $2n+4n^2(x'-\hat{x})^2$. These choices gave variances of $W_h$ that were similar for both cases and also 
for different $n$. Figure \ref{Fig:3}(b) shows how the number of data-points accessed varied with $n$ for the two methods and Figure \ref{Fig:3}(c) shows the estimates of the variance of $W_h$. As $n$ increases 
we have that the computational cost of using sub-sampling, as measured by the number of data-points accessed, remains constant. By comparison, without sub-sampling, to maintain a fixed variance for $W_h$ we need to 
increase the computational cost linearly.

Finally we looked at how the choice of $\hat{x}$ affected the performance of subsampling with control variates. We repeated the above study but setting $\hat{x}$ to be either one posterior standard-deviation or 
three posterior standard-deviations from the mode. These correspond to values in the body of the posterior and in the tail of the posterior respectively. From \ref{Fig:3} (b) and (c) we see that as $\hat{x}$ moves away 
from the posterior mode, the performance of the sub-sampler decreases both in terms of computational cost and variance. However, for a fixed variance of $W_h$ both methods have a computational cost that is constant with 
$n$, as opposed to the linearly increasing cost we have when sub-sampling is not used.


\section{Continuous-Time MCMC} \label{sec:CMCMC}

We now consider continuous-time versions of MCMC. These algorithms involve simulating a PDP process which has a given target distribution, $\pi(\x)$, as its stationary distribution. Such algorithms were originally
of interest as they are non-reversible processes. As mentioned in the introduction, there is substantial evidence that non-reversible MCMC algorithms are more efficient than standard, reversible MCMC. Intuitively this is because non-reversible
MCMC suppresses the random-walk behavour of reversible MCMC and thus can more rapidly explore the state-space. 
Furthermore it has been shown that continuous-time MCMC is suitable for using sub-sampling ideas, similar to those in Section \ref{S:SCALE}. Thus these methods are also promising for big-data applications of MCMC.


\subsection{The Continuous-time limit of MCMC} \label{sec:ctslimit}

To help build intuition for continuous-time MCMC, and to see how it links to discrete-time MCMC algorithms, we will first derive a continuous-time algorithm as a limiting form  for a simple non-reversible discrete-time
MCMC algorithm \cite[]{Gustafson1998,DiaconisHolmesNeal2000}.
This MCMC algorithm will target a joint distribution of $(\x,\bv)$, where $\bv$ can be viewed as a velocity. For our specific algorithm we will consider only velocities of a fixed, say unit, speed, 
and hence $\bv$ could equally be defined as a direction. Our MCMC will target a distribution $\pi(\x)p_u(\bv)$ where $p_u(\bv)$ will be the uniform distribution over all velocities with unit speed.

The MCMC algorithm will have two types of move. The first  involves two deterministic proposals 
\begin{itemize}
\item[(1a)] Propose a move from $(\x,\bv)$ to $(\x+h\bv,-\bv)$. Accept this with the standard Metropolis-Hastings accept probability, which simplifies to
\[
\min\left\{ 1, \frac{\pi(\x+h\bv)}{\pi(\x)}
\right\}
\]
\item[(1b)] Move from $(\x',\bv')$ to $(\x',-\bv')$.
\end{itemize}
Both the moves in (1a) and (1b) are reversible, and can be shown to satisfy detailed balance. To make step (1a) reversible we have to propose a move which flips the velocity, and hence in (1b) we flip the velocity back again. So the net affect of applying both (1a) and (1b) is that the velocity is unchanged if we accept the proposed move in step (1a) but flips if we reject the move.  
This is a standard approach in Hamiltonian Monte Carlo \cite[]{Neal:2011}. In fact this algorithm can be viewed as a type of Hamiltonian Monte Carlo move, but based on the dynamics of an approximate potential for $\x$ which is uniform (and hence the velocity is not changed other than by the flip).

Whilst this moves keeps $\pi(\x)p_u(\bv)$  invariant, it leads to a reducible Markov chain if the dimension of $\x$ is greater than 1, 
as it only proposes moves along the direction given by $\bv$. Thus we need a second type of move to produce an irreducible MCMC algorithm with the required asymptotic distribution.  
The second move we use is an update of $\bv$, from some transition kernel that has $p_u(\bv)$ as its stationary distribution. We will imagine applying $N$ transitions of type 
1 between each of these updates just of $\bv$.

Under this framework we can then consider letting $h\rightarrow 0$ while keeping $s=hN$ a constant. We will scale time so that the $i$th MCMC transition will occur at time $ih$, 
and define $(\x_t,\bv_t)$ to be the value of the state after the $i$th MCMC transition for $ih\leq t < (i+1)h$. 

Now for each move in step (1a) the rejection probability for small $h$ is
\[
\max\left\{ 
0, 1-\exp[\log\pi(\x+h\bv)-\log\pi(\x)]
\right\}=
\max\left\{ 
0, 1-\exp[\bv\cdot\nabla\log\pi(\x)h+o(h)]
\right\}=
\max\left\{ 
0, -\bv\cdot\nabla\log\pi(\x)h
\right\}+o(h),
\]
assuming that, for example, $\pi(\x)$ is twice differentiable.

Thus in our limit as $h\rightarrow0$, rejections in step (1a) will occur as events in a Poisson process of rate $\lambda(\x_t,\bv_t)=\max\{0,-\bv_t\cdot\nabla\pi(\x_t)\}$. 
The dynamics between these events will be deterministic, with $\bv_t$ being constant and $\x_t$ changing as in a constant velocity model with velocity $\bv_t$. At each event the velocity will just flip. Note that while
the process is moving to areas of higher probability density, as defined by $\pi(\x)$, the rate of the Poisson process will be 0. Thus events will only occur if the process is moving to areas of lower 
probability mass.

This limiting process is just a PDP. The dynamics of the PDP are a constant velocity model between events, with the velocity changing at event times. It is natural to consider a more general class
of PDP processes, and see what flexibility there is in choosing the distribution of the event times, and the distribution of the change of velocity at events, so that we still have a process whose marginal
stationary distribution for $\X_t$ is $\pi(\x)$.

Denote the state of our PDP by $\Z_t=(\X_t,\bV_t)$. Our class of PDPs will have the following dynamics:
\begin{itemize}
\item[(i)] {\bf The deterministic dynamics.}  For $i=1,\ldots,d$
\[
\frac{\mbox{d} x^{(i)}_t}{\mbox{d} t} = v^{(i)}_t,\mbox{ and } 
\frac{\mbox{d} v^{(i)}_t}{\mbox{d} t}=0.
\]
The solution of these dynamics is given by $(\x_{t+s},\bv_{t+s})=(\x_t+s\bv_t,\bv_t)$ for any $s>0$.
\item[(ii)] {\bf The event rate.} Events will occur at a rate, $\lambda(\mathbf{z}_t)$.
\item[(iii)] {\bf The transition distribution at events.} At an event at time $\tau$, $\x_{\tau}=\x_{\tau-}$ and
$\bv_{\tau}$ is drawn from some transition density $q(\cdot|\x_{\tau-},\bv_{\tau-})$.
\end{itemize}
The event rate and the transition density then need to be chosen so that $\pi(\x)$ is the marginal stationary distribution of the PDP.


\subsection{The Stationary Distribution of the PDP}

A necessary condition for $\pi(\x)$ to be the marginal stationary distribution of our PDP is that it is the marginal of an invariant distribution for the PDP. We will use the adjoint of the generator of
our PDP to derive a condition on both the event rate and the transition distribution at events for the PDP to have $\pi(\x)$ as the marginal of an invariant distribution.  

As above, let $\z=(\x,\bv)$. Denote the invariant distribution of our PDP by 
 $p(\z)$. We can factorise this as the product of the marginal stationary distribution for $\x$ times the conditional for $\bv$ given $\x$, and we wish to have $p(\z)=\pi(\x)p(\bv|\x)$. 
 If $\mathcal{A}^*$ is the adjoint of the generator of our PDP, as $p(\z)$ in an invariant distibution we have $\mathcal{A}^*p(\z)=0$. This gives
\begin{equation} \label{eq:stat1}
-\pi(\x)p(\bv|\x)[ \bv \cdot \nabla_\x \log \pi(\x) +\bv \cdot \nabla_\x \log p(\bv|\x) + \lambda(\z)] + \int \lambda(\x,\bv')q(\bv|\x,\bv')\pi(\x)p(\bv'|\x) \mbox{d}\bv' =0.
\end{equation}
In the above $\nabla_\x$ denotes the vector of first partial derivative with respect to the components of $\x$.

To date, all continuous-time MCMC algorithms have been designed so that under the invariant distribution $\bv$ is independent of $\x$, and thus all components of $\nabla_\x \log \pi(\bv|\x)$ will be 0. 
If we wish to design such a process we need to choose $\lambda(\x,\bv)$ and $q(\bv'|\x,\bv)$ such that (by rearranging equation \ref{eq:stat1}),
\begin{equation} \label{eq:stat}
p_v(\bv)\lambda(\x,\bv) - \int \lambda(\x,\bv')q(\bv|\x,\bv')p_v(\bv') \mbox{d}\bv'=-p_v(\bv)\bv \cdot \nabla_\x \log \pi(\x).
\end{equation}
for some distribution $p_v(\bv)$ for the velocity. The left-hand side is measuring the net probability flow out of states with velocity $\bv$, this must offset the change in probability mass for $\bV$ caused
by the deterministic dynamics, which is the term on the right-hand side.

Note that if we integrate (\ref{eq:stat}) with respect to $\bv$, the left-hand side is 0. So we get
$\mbox{E}(\bV)\cdot \nabla \log \pi(\x)=0$, 
where the expectation is with respect to the invariant distribution for the velocity. As this will need to hold for all $\x$, we can see that the invariant distribution for all components of the velocity must have 
zero mean. 

The actual processes we describe in the next section all allow velocities within some symmetrical set, and are designed so that $p_v(\bv)$ is uniform on this set. They  ensure (\ref{eq:stat}) holds through 
deterministic dynamics at events. They introduce a ``flip'' operator, $F_\x$ say, that can depend on $\x$ and which satistifies $F_\x(F_\x(\bv))=\bv$.
They then only allow transitions between pairs of velocities, $\bv$ and $\bv'$ that satisfy $\bv'=F_\x(\bv)$ and,
by definition of $F_\x$, $\bv=F_\x(\bv')$. Under this constraint on the transitions at events we get a simple set of equations that we need the event rates to satisfy. For any $\bv$, and with $\bv'=F_\x(\bv)$, it
is straightforward to show that (\ref{eq:stat}) requires
\begin{equation} \label{eq:rates}
 \lambda(\x,\bv)-\lambda(\x,\bv')=-\bv \cdot \nabla_\x \log \pi(\x).
\end{equation}
for all $\x$. Note that as this equation must also hold for $\bv'$, we immediately see that only flip operators for which $F_{\x}(\bv) \nabla_\x \log \pi(\x) = -\bv \nabla_\x \log \pi(\x)$ are allowable.
The rates only depend on the target through the term $\nabla_\x \log \pi(\x)$, which means that $\pi(\x)$ is only needed to be known up to proportionality. 
Also note
that (\ref{eq:rates}) does not uniquely define the rates. If we have a set of rates, $\lambda(\x,\bv)$ that satisfy (\ref{eq:rates}), then $\lambda(\x,\bv)+\gamma(\x,\bv)$ will also satisfy (\ref{eq:rates}) for
any positive function $\gamma(\x,\bv)$ for which $\gamma(\x,\bv)=\gamma(\x,F_\x(\bv))$. 

A natural choice of rates which satisfy (\ref{eq:rates}) are those which are smallest. This will give $\lambda(\x,\bv)=\max\{0,-\bv \cdot \nabla_\x \log \pi(\x)\}$. We will call these the {\it canonical rates}. Theoretical
justification for the  canonical rates when $d=1$ is given in \cite{Bierkens/Duncan:2016}, who show that the asymptotic variance of Monte Carlo estimators is minimised when using these rates.





\subsubsection{Different continuous-time MCMC algorithms}

We now describe a number of choices of flip operator, and the corresponding PDPs. We start with the limiting process we derived in Section \ref{sec:ctslimit}, then describe two continuous MCMC processes
that have been recently proposed. These three choices all lead to identical PDPs for a one-dimensional target, but differ in terms of how they extend to higher dimensions. Each assume the target is defined on an unbounded
domain; for extensions of these methods to bounded domains see \cite{Vollmer:2016AiStats}.
We will finish with some discussion of alternative schemes that are possible. 

{\bf Pure Reflection and Refresh}

The continuous-time limit we derived in Section \ref{sec:ctslimit} corresponds to $F_\x(\bv)=-\bv$, with the canonical rates. Such a process we call a pure reflection process. For a multi-dimensional target distribution,
this process would be reducible, as it can only explore positions $\x$ that lie on a straight-line defined by the initial velocity. As such this is an example where $\pi(\x)$ would be a marginal invariant
distribution but not the marginal stationary distribution. To overcome this we would need an additional move, which refreshes $\bv$. Such a refresh move would need to have $p_u(\bv)$ as its stationary distribution.
The times of refreshing could be either deterministic or random.

{\bf Bouncy Particle Sampler}

The Bouncy Particle Sampler of \cite{Bouchard:2016}, based on an algorithm of \cite{Peters:2012}, is an adaption of the pure reflection process, which minimises the change in
velocity at each event. It does this by defining $F_\x(\bv)$ to be 
\begin{equation} \label{eq:FlipBPS}
 F_{\x}(\bv)=\bv-2 \frac{\bv\cdot \nabla \log \pi(\x)}{\nabla \log \pi(\x)\cdot \nabla \log\pi(\x)}\nabla \log \pi(\x).
\end{equation}
This flips the component of $\bv$ that is in the direction of $\nabla \log \pi(\x)$ but leaves the components of $\bv$ that are orthogonal to $\nabla \log \pi(\x)$ unchanged. They again use the canonical rates. As with
the pure reflection process this means that events only occur if the PDP is moving to areas of lower probability mass according to $\pi(\x)$. 

The original sampler of \cite{Peters:2012} just implement that above sampler. However \cite{Bouchard:2016} shows that, for some targets, such a sampler can be reducible. The Bouncy Particle Sampler introduces
a refresh step, which occurs at events of an independent Poisson process of constant rate. Furthermore they show that for any non-zero rate of this refresh process, the resulting sampling will have $\pi(\x)$ as 
its asymptotic distribution. 


{\bf Zig-Zag Sampler}

The Zig-Zag sampler \cite[]{Bierkens/Roberts:2015,Bierkens/Fearnhead/Roberts:2016} considers a discrete set of velocities. If $\x$ is $d$-dimensional, then $\bv=\sum_{i=1}^d \theta_i \be_i$, where each $\theta_i\in\{-1,1\}$
and $\be_{1},\ldots,\be_{d}$ are a set of orthongal basis vectors for $\R^d$. The invariant distribution for $\bv$ is defined as the uniform distribution over this set of $2^d$ possible values. 

The Zig-Zag sampler can be viewed as having $d$-distinct event types, each with its own rate, and each with its own deterministic change to the velocity. The $i$th event will have a flip, $F^{(i)}$, that switches $\theta_i$ to 
$-\theta_i$, but keeps the velocity in the other $d-1$ directions unchanged. If we denote $\lambda_i(\x,\bv)$ to be the rate of events of type $i$, then this corresponds to our general formulation of a PDP but with
$\lambda(\x,\bv)=\sum_{i=1}^d \lambda_i(\x,\bv)$, and with the transition distribution at an event being a discrete distribution over the $d$ transitions that correspond to the $d$ different flips. Flip $i$ occurs 
with probability $\lambda_i(\x,\bv)/\lambda(\x,\bv)$. Subsituting this into (\ref{eq:stat}) shows that we need to choose $\lambda_i(\x,\bv)$ so that
\[
 \sum_{i=1}^d \left\{ \lambda_i(\x,\bv) - \lambda_i(\x,F^{(i)}(\bv)) \right\} = -\sum_{i=1}^d \theta_i \frac{\partial\log \pi(\x)}{\partial x^{(i)}}.
\]
Here we have assumed that $x^{(i)}$ is the component of $\x$ in direction $\be_i$. This can be achieved if we choose the rates such that
\[
 \lambda_i(\x,\bv) - \lambda_i(\x,F^{(i)}(\bv)) = -\theta_i \frac{\partial\log \pi(\x)}{\partial \x^{(i)}}.
\]
As above, this does not uniquely define the rates, only the difference between rates for velocities that differ in terms of their component in the $\be_i$ direction.

It is a challenging goal to show that the Zig-Zag process is ergodic in full generality. So far it is established rigorously in \cite{Bierkens/Fearnhead/Roberts:2016} that the Zig-Zag process is ergodic in any of the following cases: (i) one-dimensional target distributions, (ii) factorized target distributions, and (iii) switching rates that are positive everywhere (which can be obtained by adding a constant $\varepsilon > 0$ to the canonical switching rates). Experiments suggest that ergodicity holds in much more generality.

Note that the above argument easily generalises to allowing velocities of the form $\bv=\sum_{i=1}^m \theta_i \be_i$, where the $\be_i$ are not constrained to be orthogonal, and we can even allow $m>d$ directions. Whether
there are advantages in using such a set of possible velocities is not clear.

{\bf Alternatives}

There is substantial extra flexibility in choosing the event rates and the type of transition at events beyond the three examples we have detailed. For example we could consider transitions at an event that
does not depend on the current velocity. If we allow $\bv$ to be any unit vector, then
it is straightforward to show that choosing $\lambda(\x,\bv)=\max\{0,-\bv \cdot \nabla_\x \log \pi(\x)\}$, and, at each event, sampling a new velocity from the distribution
\[
 q(\bv'|\x,\bv)\propto \max\{0,\bv' \cdot \nabla_\x \log \pi(\x)\},
\]
will lead to a PDP with invariant distribution that has $\pi(\x)$ as its marginal. 

More substantial alternatives are also possible. For example, we could consider processes which allow
the invariant distribution of $\bV$ to depend on $\x$ -- something that \cite{Girolami:2011} has shown to be beneficial for Hamiltonian Monte Carlo methods. For a proposed distribution $\pi(\bv|\x)$ we would then
need to find a set of event rates and transitions that satisfy (\ref{eq:stat1}). 


\subsection{Simulation and Use of Skeletons for Continuous MCMC} \label{sec:skeletons}

So far we have described a number of different PDPs that will have $\pi(\x)$ as their marginal invariant distribution. For these to be useful in practice, we need to be able to simulate them efficiently. How to do
this in practice will depend on the form of $\pi(\x)$, but is likely to use the ideas briefly described at the end of Section \ref{sec:Sim}. For further detail see the discussion of this, and suggestions, in
\cite{Bouchard:2016} and \cite{Bierkens/Fearnhead/Roberts:2016}.

The output of simulating a PDP will be a set of event times and the values of the state at those event times. We wish to use this output to obtain Monte Carlo estimates of expectations of functions of $\X$, where
$\X$ is distributed according to $\pi(\x)$. Assume we have simulated the PDP for some time-interval $T$. We will discard the value of the process in some burn-in period of length $t_b$. Assume there were $N$
events in the time-interval $[t_b,T]$. Denote these as $\tau_i$ for $i=1,\ldots,N$, and let $\tau_0=t_b$ and $\tau_{N+1}=T$. 

There are two approaches to obtain a Monte Carlo estimate of $\int \pi(\x) g(\x) \mbox{d}\x$ for some function $g(\x)$ of interest. The first is to calculate the average of this function along the path of the PDP:
\[
\frac{1}{\tau_{N+1}-\tau_0} \sum_{i=0}^N \int_{0}^{\tau_{i+1}-\tau_{i}} g(\x_{\tau_i}+s\bv_{\tau_i}) \mbox{d}s.
\]
Here each integral corresponds to the integrals of $g(\x_t)$ for $t$ in $[\tau_i,\tau_{i+1}]$, and uses the fact that for such a $t$, $\x_t=\x_{\tau_i}+(t-\tau_i)\bv_{\tau_i}$. 

The above approach is difficult if the integrals are not easy to evaluate. In this case we can resort to a standard Monte Carlo approximation. Choose an interger $M>0$, define $h=(\tau_{N+1}-\tau_0)/M$ and then
use the Monte Carlo estimator
\[
 \frac{1}{M} \sum_{j=1}^M g(\x_{\tau_0+jh}),
\]
where, as above, we can trivially calculate $\x_{\tau_0+jh}$ using the set of event times and the values of the PDP at those event times. 


\subsection{Exact Approximation versions and Subsampling}

Exact approximate algorithms \cite[]{Andrieu/Roberts:2009} are MCMC algorithms that use estimators of the target distribution within the accept-reject step. If implemented correctly, and if these estimators are both positive and unbiased, then
it can be shown that the resulting MCMC algorithms are exact: in the sense they still have the target distribution as their stationary distribution. It turns out that exact approximate versions of 
the continuous-time MCMC algorithms detailed in the previous section are also possible.

\subsubsection{Exact Approximation for Pure Reflection and Zig Zag}

For concreteness and ease of presentation we will consider an exact approximate version of the Pure Reflection process. Though the ideas we detail extend trivially to the Zig-Zag sampler \cite[and see][for more details 
of an exact approximate version of Zig-Zag]{Bierkens/Fearnhead/Roberts:2016}. 

For the Pure Reflection process
the requirement on the rates of events is that for any
velocity $\bv$
\[
 \lambda(\x,\bv)-\lambda(\x,-\bv)= - \bv \cdot \nabla \log \pi(\x).
\]
For a given choice of rates, such as the canonical rates $\lambda(\x,\bv)=\max\{0, - \bv \cdot \nabla \log \pi(\x)\}$, we would often use thinning to simulate the event times (see Section \ref{sec:Sim}). Thus if
our current state is $(\x_t,\bv_t)$  we would introduce a bound on the event rate for $s>0$
\[
 \tilde{\lambda}^+(s)\geq \lambda(\x_t+s\bv_t,\bv_t),
\]
simulate potential events at rate $\tilde{\lambda}^+(s)$ and accept them with probability $\lambda(\x+s\bv,\bv)/\tilde{\lambda}^+(s)$. The time until the next event is just the time until the first accepted event.

Now assume we have a estimator of $-\nabla \log \pi(\x)$, which we will denote $\U(\x)$. This estimator is a random variable, and examples of how it could be constructed are given below. We further introduce a random
rate function
\[
 \hat{\lambda}(\x,\bv)=\max\{0, \bv \cdot \U(\x)\}.
\]
This is just the canonical event rate, but replacing $-\nabla \log \pi(\x)$ with its unbiased estimator. The idea of an exact-approximate version is to simulate events using thinning, 
with a bound on the event rate that satisfies
\[
 \tilde{\lambda}^+(s)\geq \hat{\lambda}(\x_t+s\bv_t,\bv_t)
\]
almost surely, and where we accept points using the random acceptance probability $ \hat{\lambda}(\x_t+s\bv_t,\bv_t)/\tilde{\lambda}^+(s)$. 

As the overall acceptance probability will be the expectation of the random acceptance probability, it is straightforward to show that simulating events in this way is equivalent to simulating events at a rate 
\begin{equation} \label{eq:exprate}
\lambda(\x,\bv)=\mbox{E}\left( \max\left\{0,\bv\cdot \U(\x) \right\}\right),
\end{equation}
where expectation is with respect to the random variable $\U(\x)$. Furthermore if we now calculate the difference in rates, $\lambda(\x,\bv)-\lambda(\x,-\bv)$, we have
\begin{eqnarray*}
 \lambda(\x,\bv)-\lambda(\x,-\bv)&=&\mbox{E}\left( \max\left\{0,\bv\cdot \U(\x) \right\}\right)-\mbox{E}\left( \max\left\{0,-\bv\cdot \U(\x) \right\}\right) \\
 &=& \mbox{E}\left( \max\left\{0,\bv\cdot \U(\x) \right\}- \max\left\{0,-\bv\cdot \U(\x) \right\}\right) \\
  &=& \mbox{E}\left( \max\left\{0,\bv\cdot \U(\x) \right\}+ \min\left\{0,\bv\cdot \U(\x) \right\}\right) 
=\mbox{E}\left( \bv\cdot \U(\x) \right).
\end{eqnarray*}
Thus provided $\U(\x)$ is an unbiased estimator of $-\nabla \log \pi(\x)$, the resulting process will have $\pi(\x)$ as its marginal invariant distribution.

Note that using an unbiased estimator of $-\nabla \log \pi(\x)$ does not come without cost, as the resulting process will, in general, be less efficient. This loss of efficiency comes first as the rates that are used
for events, $\mbox{E}( \max\{0,\bv\cdot \U(\x)\})$, will, in general, be larger than the canonical rates. The only exception being if $\U(\x)$ has the same sign as $-\nabla \log \pi(\x)$ with probability one. 
Using larger rates appears to reduce the rate of mixing of the process  \cite[see][]{Bierkens/Duncan:2016,Bierkens/Fearnhead/Roberts:2016}. A related issue is that the bound on the rates, used when implementing
thinning, will also tend to be larger. This will increase the cost of simulating the process. Intuitively we would expect both these losses of efficiency to increase as the variability of our estimator increases.


\subsubsection{Exact Approximation for the Bouncy Particle Sampler}

The idea of an exact approximation version of the Bouncy Particle Sampler is slightly more complicated, due to the fact that the flip operator used also depends on $\nabla \log \pi(\x)$. To implement such an
exact approximation version we need to use the same estimate of $\nabla \log \pi(\x)$ in the flip operator as was used in deciding to accept the event. So, using the notation of the previous section, at a potential
event at time $s$, simulated from a Poisson process with rate $\tilde{\lambda}^+(s)$ we will now:
\begin{itemize}
 \item[(1)] Simulate $\bu$, a realisation of $\U(\x_t+s\bv_t)$.
 \item[(2)] Accept the event with probability 
 \[
  \frac{\hat{\lambda}(\x_t+s\bv_t,\bv_t)}{\tilde{\lambda}^+(s)}=\frac{ \max\{ 0, \bu \cdot \bv_t \} }{\tilde{\lambda}^+(s)}
 \]
 \item[(3)] If we accept the potential event, this corresponds to an event at time $t+s$, with new state being $\x_{t+s}=\x_{t}+s\bv_t$ and
 \[
  \bv_{t+s}= \bv_t -2 \frac{\bv_t\cdot \bu}{\bu \cdot \bu}\bu.
 \]
 \end{itemize}
Note that the same realisation, $\bu$, is used in both steps (2) and (3). 

This leads to an algorithm where the transition at an event is random. We will assume that $\U$ is a discrete random variable, which is consistent with the use of sub-sampling discussed in Section \ref{sec:Subsample}.
It is straightforward to show that the resulting PDP has events occurring at rate (\ref{eq:exprate}) as before, but with a transition 
probability mass function at an event that is
\[
 q(\bv'|\x,\bv)= p(\bu|\x)\frac{\max\{ 0, \bu \cdot \bv \}}{ \mbox{E}\left( \max\left\{0,\bv\cdot \U(\x) \right\}\right)} 
\]
where $p(\bu|\x)$ is the probability mass of simulating $\bu$, and $\bv'=\bv-2\bu(\bv\cdot\bu)/(\bu\cdot\bu)$. Now substituting these values into the equation for the stationary distribution of the PDP (\ref{eq:stat1}), it
is simple to show that the resulting process will have an invariant distribution $\pi(\x)p_v(\bv)$, where $p_v(\bv)$ is uniform on the set of velocities of fixed speed.


\subsubsection{Use of Subsampling} \label{sec:Subsample}

An example of an exact-approximate version of these continuous-time MCMC algorithms arises if we use sub-sampling of data points at each iteration when performing 
Bayesian inference in a big-data setting. As in Section \ref{S:SCALE}, we will consider a target density
of the form
 \[
 \pi(\x)\propto \prod_{i=1}^n \pi_i(\x).
\]
In this case, the simplest unbiased estimator of $-\nabla \log \pi(\x)$ is just
\begin{equation} \label{eq:simp_subsamp}
 -n \nabla \log \pi_I(\x),
\end{equation}
where $I$ is drawn uniformly from ${1,\ldots,n}$. However, to increase the efficiency of sub-sampling methods we would want to try and minimise the variance of our estimator, for example by using
control variates. As in Section \ref{S:SCALE}, one approach is to have a pre-processing step that finds $\hat{\x}$, a value of $\x$ that is near the posterior mode. We then calculate and store
$\nabla \log \pi(\hat{\x})$, and use the following estimator
\begin{equation} \label{eq:cv}
 -\nabla \log \pi(\hat{\x}) + n\left(\nabla \log \pi_I(\hat{\x})- \nabla \log \pi_I(\x) \right).
\end{equation}
By a similar argument to that in Section \ref{S:SCALE}, if $\x$ is within a distance of $O(\sqrt{n})$ of $\hat{\x}$ and if $\pi(\x)$ is sufficiently smooth we would expect the variance of this estimator
to only increase linearly with $n$. By comparison the variance of the simple estimator (\ref{eq:simp_subsamp}) will increase like $O(n^2)$.

\subsection{Example: Mixture Model}

To demonstrate some of the properties of these continuous-time MCMC algorithms we apply them to the simple mixture model of Section \ref{sec:MM1}. This involves inference for a univariate parameter, and for this
case each of the three versions of continuous-time MCMC introduced earlier are equivalent. Furthermore we do not need to introduce any refreshing of the velocity to ensure ergodicity 
\cite[]{Bierkens/Duncan:2016}. Our aim is to show how the continuous-time MCMC algorithms can be implemented, how the choice of the bounding process that simulates potential event times can effect efficiency, and give some insight into how and when
the subsampling ideas can lead to gains in efficiency. 

We will first look at implementing continuous-time MCMC using a global bound on the event rates.
For this model, if we write $\pi(x)\propto \prod_{i=1}^n \pi_i(x)$, where $\pi_i$ is the likelihood of the $i$th observation times the $1/n$th root of the prior, then
\[
 \log \pi_i(x) = \log\left(\frac{p}{10}\exp\left\{-\frac{1}{200}y_i^2 \right\}+(1-p)\exp\left\{ -\frac{1}{2}(x-y_i)^2 \right\} \right) - \frac{1}{8} x^2.
\]
As in Section \ref{sec:MM1} we will fix $p=0.95$ in the simulations we present.
We can bound $|\nabla \log \pi_i(x)|$ for each $i$, and this bound increases with $|y_i|$. If we let $j$ be the observation with largest absoluate value,  then the simplest global bound on the event rates will be
\begin{equation} \label{eq:lambda1}
 \lambda^+ = n \max_x |\nabla \log \pi_j(x)|.
\end{equation}

We can then simulate the path of our continous-time MCMC algorithm by iterating the following steps. Assuming we are currently at time $t$ with state $(x_t,v_t)$:
\begin{itemize}
 \item[(1)] Simulate the time until the next putative event, $s$, a realisation of an exponential distribution with rate $\lambda^+$.
 \item[(2)] Calculate $x_{t+s}=x_t+sv_t$, and the actual rate of an event at position $x_{t+s}$:
 \[
  \lambda(x_{t+s},v_t)=\max\{0, -v_t \nabla \log \pi(x_{t+s}) \}.
 \]
\item[(3)] With probability $\lambda(x_{t+s},v_t)/\lambda^+$ switch the sign of the velocity, $v_{t+s}=-v_t$ and store the value $(x_{t+s},v_{t+s})$. Otherise $v_{t+s}=v_t$. 
\end{itemize}
This simulates using the canonical rates. To use the simplest version of sub-sampling we just replace the calculation of the actual rate in step (2) by
\[
 \lambda(x_{s+t},v_t)=n\max\{0, -v_t \nabla \log \pi_I(x_{s+t}) \},
\]
for $I$ sampled uniformly from $\{1,\ldots,n\}$. Note that our choice of $\lambda^+$ can still be used with sub-sampling, as it bounds the above rate for all $I$ and $x_{s+t}$.

The above algorithm has similarities to one iteration of a standard MCMC algorithm. Steps (1) and (2) can be viewed in terms of simulating a new state, and step (3) is a form of accept-reject step. However 
there are fundamental differences. Firstly, the probability in step (3) depends on the target through $\nabla \log \pi(x)$, as compared to the acceptance probability in MCMC which depend on $\pi(x)$.
Secondly, the algorithm moves from $x_s$ to $x_{s+t}$ regardless of the outcome in step (3). Step (3) only affects the velocity component. Finally, as mentioned in Section \ref{sec:skeletons} 
the use of the output is different. For continuous-time MCMC we have to take averages with respect to the continuous-time path, or with respect the value of the process at equally-spaced time-points. By comparison
MCMC would average with respect to the value of the chain at the end of each iteration.

\begin{figure}
 \centering
 \includegraphics[scale=0.6]{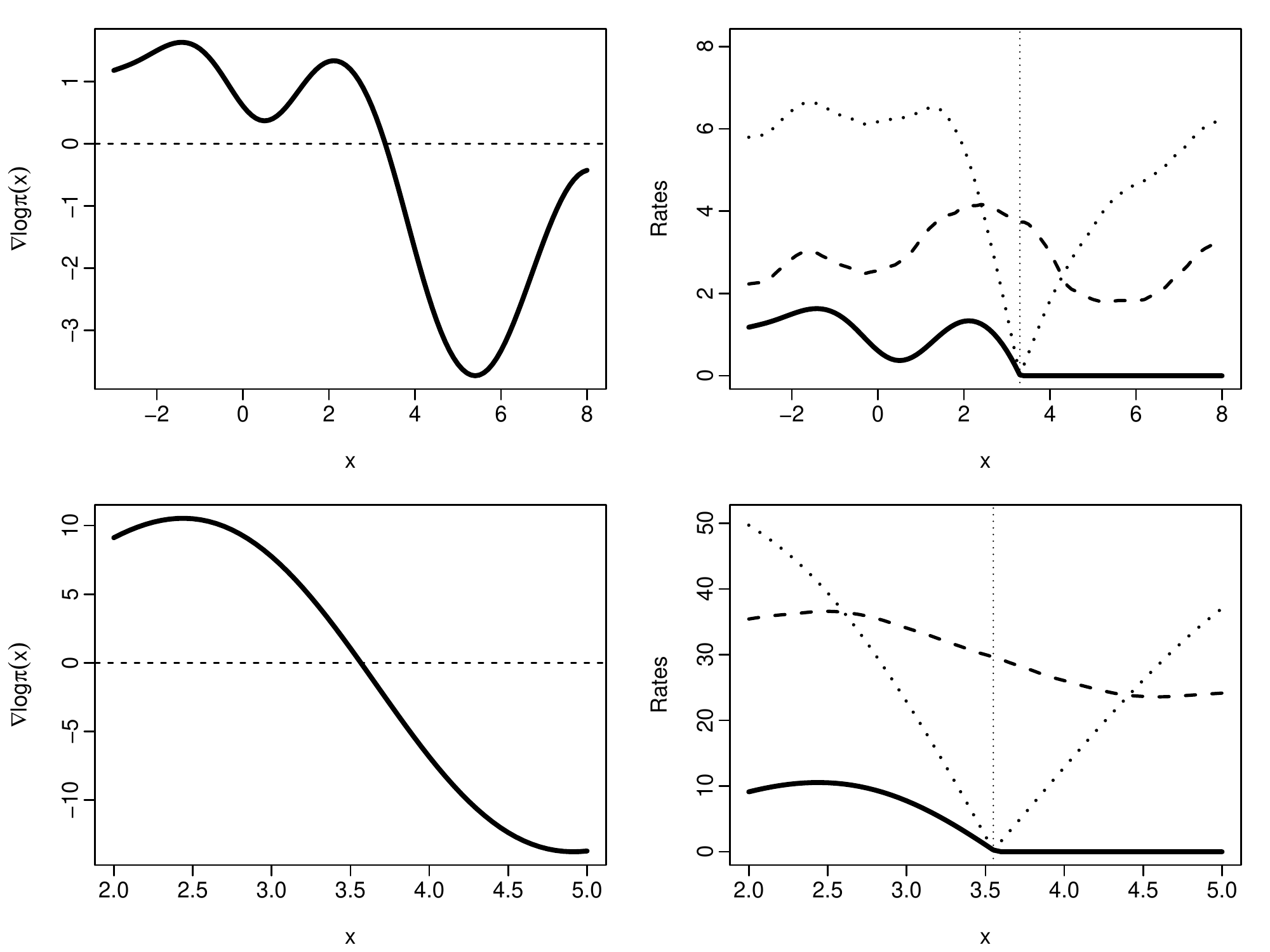}
 \caption{\label{Fig:Lambdas} 
 Plots of $\nabla \log \pi(x)$ for the mixture example (left-hand column), and rates at which the continuous-time MCMC algorithm will switch from a negative to a positive velocity (right-hand column). For the latter 
 plots we show rates for the canonical process (full lines), simple sub-sampling (dashed lines) and sub-sampling with control variates (dotted lines). The vertical dotted line shows the value of $\hat{x}$.
 Top row if for 150 data points, and the bottom row for 1,500 data points.  Plots are restricted to areas of non-negligible posterior mass.
 }
 \end{figure}

We now turn to how the use of sub-sampling impacts on the efficiency of continuous-time MCMC. For our above implementation the average number of iterations needed to simulate a path over a time-interval of length 
$T$ will be $T/\lambda^+$ for both the canonical and sub-sampling versions. Sub-sampling will involve a smaller cost in step (2) as the rate depends on just a single data point rather than all $n$ data point. However
this computational saving comes at the cost of an overall increased rate of switching velocity. This is shown in Figure \ref{Fig:Lambdas}, where we give examples of $\nabla \log \pi(x)$ for two simulated data sets, 
of size $n=150$ and $n=1,500$ respectively, and each simulated with the true value of $x=4$. We also show the canonical rate of switching from a negative to a positive velocity, and the expected rate of switching 
when we use sub-sampling. The canonical implementation has 
uniformly lower rates. 

\begin{figure}
 \centering
 \includegraphics[scale=0.6]{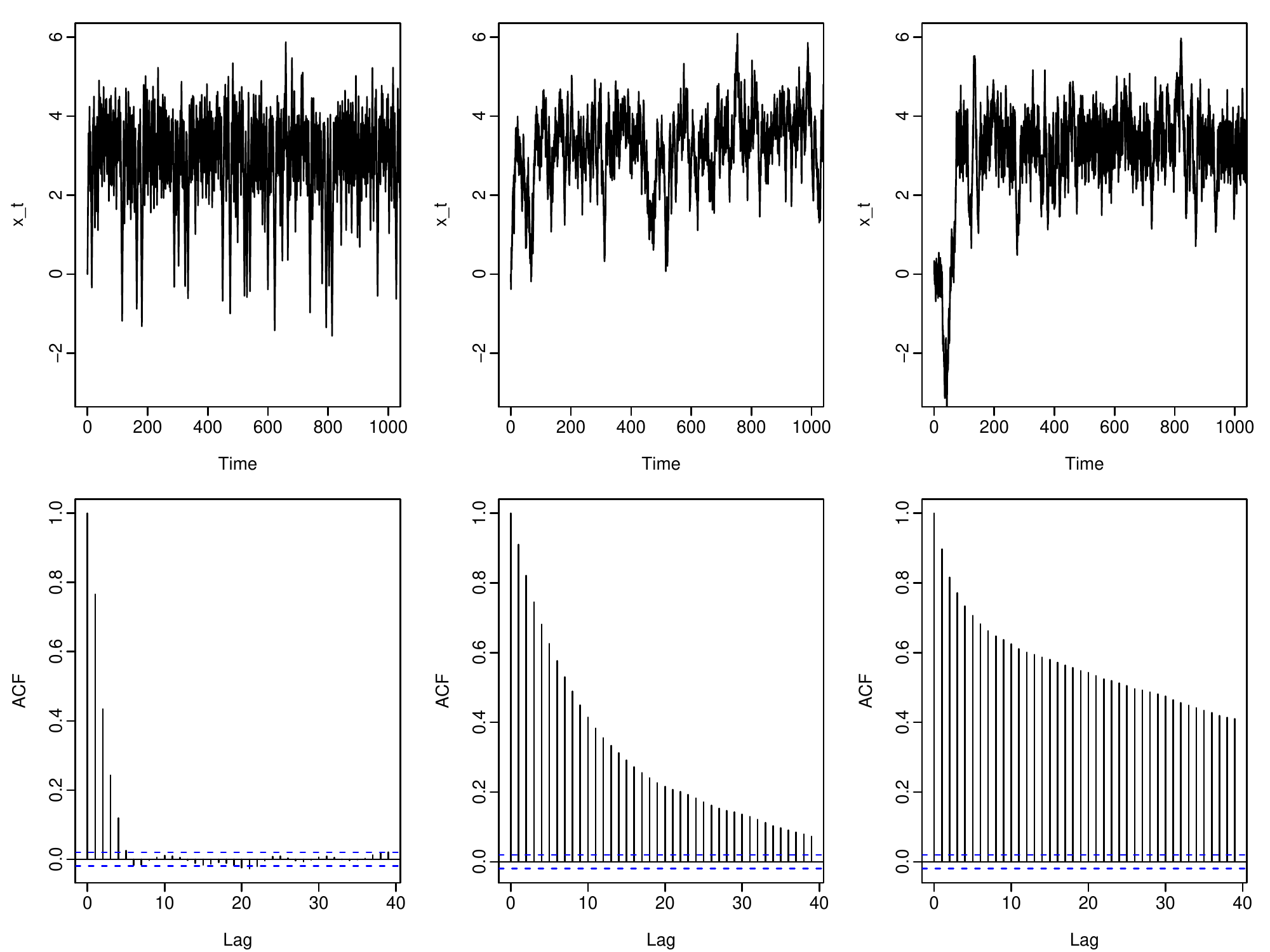}
 \caption{\label{Fig:ZigZag} 
 Trace plots (top row) and autocorrelation plots (bottom row) for three implementations of continuous-time MCMC: canonical process (left-hand column); simple sub-sampling (middle column) and subsampling with control
 variates (right-hand column). Auto correlation plots are values sampled every unit time-step from the continuous sample paths.
 }
 \end{figure}

The impact of these different rates can be seen in Figure \ref{Fig:ZigZag}, where we show trace autocorrelation plots for analysing the data set with $n=150$. Using subsampling leads to paths of the sampler that
switch velocity substantially more frequently. As a result, the canonical implementation is more efficient in terms of suppressing random walk behaviour, and this is seen in terms of better mixing and lower
autocorrelation. The autocorrelation plots suggest we need to run the sub-sampling version for roughly 5 times as long to obtain the same accuracy as using the canonical implementation.

We can improve the computational efficiency of both these implementations of continuous-time MCMC through using a lower bounding rate, $\lambda^+$. The possibility for lowering $\lambda^+$ is greater, however, 
for the canonical version, and this is a second advantage it has over using sub-sampling. For example, with an additional pre-processing cost, we could choose
\begin{equation} \label{eq:lambda2}
  \lambda^+ = \sum_{i=1}^n \max_x |\nabla \log \pi_i(x)|.
\end{equation}
Such a choice can be used with sub-sampling if our estimate of the rate uses non-uniform sampling of data points:
\[
 \lambda(x,v) = \lambda^+ \max\left\{0,-v \frac{\nabla \log \pi_I(x)}{\max_x |\nabla \log \pi_I(x)| } \right\},
\]
where we sample $I$ from ${1,\ldots,n}$, with value $i$ having probability proportion to $\max_x |\nabla \log \pi_i(x)|$. For the canonical implementation we can reduce the rate further if we are able to use the actual
maximum of $|\nabla \log \pi(x)|$, but such a choice is not valid with sub-sampling. For our example, using (\ref{eq:lambda2}) rather than (\ref{eq:lambda1}) will reduce the number of iterations required by a factor of
5.3. If we used the actual maximum of $|\nabla \log \pi(x)|$ for the canonical version, we would reduce the number of iterations by a factor of nearly 30 when compared to using (\ref{eq:lambda1}).
Note that for all these options for choosing $\lambda^+$, the underlying stochastic process we are simulating is unchanged -- it is just the efficiency of the simulation algorithm that is affected.

If we compare the best implementation of continous-time MCMC with subsampling, using  (\ref{eq:lambda2}), to the best version of the canonical implementation we get that for the same accuracy the
we would need just over 25 times as many iterations using subsampling. Each iteration would be quicker, however, as it would need access just one, out of 150, data-points. 


To see any substantial gains from using subsampling, we need to have a lower variance estimator of $\nabla \log \pi(x)$, using, for example, control variates (\ref{eq:cv}). To implement this we need to upper bound
our estimator of the rate. This is possible for this application as the absoluate value of the second derivate of $\log \pi_i(x)$ is bounded. Assume we can find a bound, $C$ say, then we use a bounding rate that
depends of the form
\begin{equation} \label{eq:lambda_cv}
\lambda^+(x) = |\nabla \log \pi(\hat{x})| + nC|x-\hat{x}|.
\end{equation}

To implement the resulting algorithm we can again iterate the three steps given above. The only changes are that in step (1) we need to simulate the inter-event time from a point process with rate
\[
 \tilde{\lambda}^+(s)=|\nabla \log \pi(\hat{x})| + nC|x_t+v_ts-\hat{x}|,
\]
and in step (3) the probability of switching the velocity is
\[
 \frac{\max\{0,-v[\nabla \log \pi(\hat{x})+n(\nabla \log \pi_I(x_{t+s})-\nabla \log \pi_I(\hat{x}) ) ]\}}{|\nabla \log \pi(\hat{x})| + nC|x_{t+s}-\hat{x}|}.
\]

We can get some insight into the advantage of using control variates by calculating the expected rate of switching the velocity for the resulting algorithm and comparing with this rate for
the other two implementation. This comparison is shown in Figure \ref{Fig:Lambdas}. We see a much lower rate of switching when we use control variates if $x$ is close to $\hat{x}$ as compared to
the simple sub-sampling approach. However the rate is actually larger if $x$ is far from $\hat{x}$. Thus we see the importance of  $\hat{x}$  being close to the mode of the posterior. This picture
is the same for both small, $n=150$, and larger, $n=1,500$, data sets. However for larger data sets the posterior mass close to the posterior mode increases. As such the amount of time that algorithm
will be in regions where using control variates is better will increase as we analyse larger data sets.

In Figure \ref{Fig:ZigZag} we see output from the algorithm using control variates for $n=150$. For such a small sample size, there appears to be little advantage in using control variates. The mixing in the tails 
is poor, due to the large variability of our esimators of the switching rate when $x$ is not close to $\hat{x}$. Note that we can avoid this issue by using a hybrid scheme that estimates the rates
using control variates when $|x-\hat{x}|$ is small, and uses simple sub-sampling otherwise.

We see advantages from using control variates as we analyse larger data sets. A comparison of our three implementations of continuous-time MCMC is shown in Table \ref{tab:1}. Firstly note that for the canonical 
implementation, the amount of time we need to run the continuous-time MCMC for decreases with sample size. This is as described in the scaling limits discussed in \cite{Bierkens/Fearnhead/Roberts:2016}.
The intuition is that for larger $n$ the posterior is more concentrated, and thus the underlying PDP process needs less time to explore the posterior. This property is also seen if we use subsampling
with control variates. Without control variates, the actual switching rates of the underlying PDP increase quickly with $n$ which slows down the mixing of the algorithm, and the amount of time
we need to simulate the underlying PDP for does not change much with $n$. 

The actual computational cost also depends on the number of iterations of the algorithm per unit-time interval. This increases with $n$ for all implementations. However the overall computational cost,
as measured by iterations per ESS remains roughly constant when we use control variates. As the computation cost per iterations is $O(1)$ we see evidence that this algorithm has a computational cost
that does not increase with $n$. By comparison the number of iterations per ESS appears to increase roughly linearly with $n$ if we use subsampling without control variates. For the canonical implementation, even
using the best possible global bound on the event rate, we have the number of iterations per ESS remaining constant but the computational cost per iterations is $O(n)$. Thus its overall
computational cost will increase linearly with $n$.

\begin{table}
\begin{center}
\begin{tabular}{|lr|cc|c|c|}
 \hline
& & Canonical & Canonical & Subsampling & Control Variate \\
& Bounding Rate             & (\ref{eq:lambda2}) & $\min |\nabla \log \pi(x)|$ & (\ref{eq:lambda2}) & (\ref{eq:lambda_cv}) \\ \hline 
$n=150$&$t$ per ESS                &   3.3              & 3.3 			      &14		&22 \\   
&Iterations per unit time  &  57			&11			      &57		&210 \\   
&Iteration per ESS         & 190  			&36			      &800		& 4,600\\ \hline   
 $n=1,500$&$t$ per ESS       &   0.43              & 0.43 			      &8.6		& 2.1 \\   
&Iterations per unit time  &  570			&15			      &570		&1,000 \\   
&Iteration per ESS         &  245 			&6.4			      &4,900		&2,100 \\ \hline   
 $n=15,000$&$t$ per ESS       &   0.13              & 0.13 			      &9.1		&0.91 \\   
&Iterations per unit time  &  5,600			&100			      &5,600		&3,800\\   
&Iteration per ESS         &  730 			&13			      &51,000		&3,500\\ \hline   
 \end{tabular}
 \caption{\label{tab:1}
 Comparison of different implementations of continuous-time MCMC: canonical, subsampling, and subsampling with control variates; and
 how they vary as sample size, $n$, increases. Both canonical and subsampling use a global bound on the event rate to simulate
 possible events, we give results for canonical using both (\ref{eq:lambda2}) and $\min |\nabla \log \pi(x)|$ as this bound. 
 We give estimates of the time-length the MCMC algorithm needs to be run for each
 effective sample size (ESS) and the average number of iterations per unit time-interval. The product of these is then the number
 of iterations needed per ESS. Remember that the cost per iteration is $O(n)$ for canonical, and $O(1)$ for subsampling and
 subsampling with control variates. }
\end{center}
 \end{table}
 

\section{Discussion}

We have shown how piecewise deterministic processes can be used to derive continuous-time versions of sequential Monte Carlo and MCMC algorithms. These algorithms are fundamentally different from more standard
discrete-time versions. Currently only a few specific algorithms, from a much wider class of possibilities, have been suggested. Whilst we have suggested a few extensions of existing methods, these just
touch the surface of the range of developments that are possible. Whilst not discussed here, the continuous-time SMC methods seem particularly well-suited for implementation on a distributed computing architecture,
as evolution of particles can be carried out in parallel. As well as such potential methodological developments, there are a wide-range of open theoretical questions. For example, can we get results on how
well continuous-time MCMC mixes? For such results to be practically meaningful they would need to account for the computational cost of simulating the underlying PDP, as opposed to just measuring the
mixing properties of the PDP itself. Can we characterise the situations where continuous-time MCMC is more efficient than its discrete-time counterpart? Or understand which versions of continuous-time MCMC are most 
efficient, and when?

We have also shown how subsampling ideas, where we approximate the gradient of the log-posterior using a small sample of data points, can be used with these continuous-time methods. Unlike other
subsampling approaches, the methods still remain ``exact'', in the sense that they still target the true posterior. Subsampling reduces the computational cost per iteration but does lead to a 
increase in Monte Carlo error for a fixed number of iterations. In the examples we have considered, it is only when using control variate ideas to reduce the variance of our sub-sampling estimator of
the gradient of the log posterior, that we see any overall gain in efficiency of the algorithm. Furthermore, when using suitable control variates, it appears possible to obtain algorithms whose computational cost
per effective sample size increases sub-linearly with the number of data points. This adds to existing evidence of the importance of using control variates if we wish to have some form of super-efficiency for
big data problems \cite[]{Bardenet:2015}. 

{\bf Acknowledgements} The authors think the Engineering and Physical Sciences Research Council for support through grants EP/K014463/1 (i-Like) and EP/D002060/1 (CRiSM).

\bibliography{refs,zigzag,joris}

\end{document}